\documentclass[lettersize,journal]{IEEEtran}
\usepackage{amsmath,amsfonts}
\usepackage{algorithmic}
\usepackage{algorithm}
\usepackage{array}
\usepackage{bm}
\usepackage{textcomp}
\usepackage{stfloats}
\usepackage{url}
\usepackage{verbatim}
\usepackage{graphicx}
\usepackage{cite}
\usepackage{xcolor}
\usepackage[font=small,labelfont=bf]{caption}
\usepackage[font=small,labelfont=bf]{subcaption}
\usepackage{makecell}
\usepackage{multicol}
\usepackage{multirow}
\usepackage{booktabs}
\usepackage{mathtools}
\usepackage{enumitem}
\usepackage{booktabs}
\usepackage{tabularx}

\newcommand{\dma}{\mathsf{dma}}
\newcommand{\ps}{p}
\newcommand{\Pps}{N_\mathsf{ps}}
\newcommand{\Nsub}{N_\mathsf{sub}}
\newcommand{\Nrf}{N_\mathsf{rf}}
\newcommand{\Nps}{N_\mathsf{ps}}
\newcommand{\rmf}{\mathrm{f}}
\newcommand{\ana}{\mathsf{ana}}

\newcommand{\rf}{r}

\newcommand{\transpose}{\mathrm{T}}
\newcommand{\btg}{\beta_\mathsf{g}}

\newcommand{\Pink}{P_{\mathsf{in},k}}
\newcommand{\be}{\mathbf{e}}
\newcommand{\Nx}{N_\mathsf{x}}
\newcommand{\Ny}{N_\mathsf{y}}
\newcommand{\Gre}{G_\mathsf{realized}}
\newcommand{\Pcons}{P_\mathsf{cons}}
\newcommand{\coupling}{\nu}

\newcommand{\Fnorm}{\mathsf{F}}

\newcommand{\secref}[1]{{Section}~\ref{#1}}
\newcommand{\figref}[1]{{Fig.}~\ref{#1}}

\newcommand{\bydef}{\triangleq}

\def\bydef{:=}

\def\bb0{{\mathbb{0}}}

\def\bydef{:=}

\def\bb{{\mathbf{b}}}

\def\bd{{\mathbf{d}}}

\def\bff{{\mathbf{f}}}

\def\bn{{\mathbf{n}}}

\def\bp{{\mathbf{p}}}
\def\bq{{\mathbf{q}}}

\def\bv{{\mathbf{v}}}

\def\bx{{\mathbf{x}}}
\def\by{{\mathbf{y}}}
\def\bz{{\mathbf{z}}}
\def\b0{{\mathbf{0}}}
\def\bA{{\mathbf{A}}}

\def\bF{{\mathbf{F}}}
\def\bG{{\mathbf{G}}}
\def\bH{{\mathbf{H}}}
\def\bI{{\mathbf{I}}}

\def\bP{{\mathbf{P}}}
\def\bQ{{\mathbf{Q}}}

\def\bT{{\mathbf{T}}}

\def\bW{{\mathbf{W}}}

\def\bZ{{\mathbf{Z}}}

\def\bbC{{\mathbb{C}}}

\def\bbR{{\mathbb{R}}}

\def\cA{\mathcal{A}}

\def\cF{\mathcal{F}}

\def\cH{\mathcal{H}}
\def\cI{\mathcal{I}}

\def\cR{\mathcal{R}}
\def\cS{\mathcal{S}}

\def\sfA{\mathsf{A}}

\def\sfH{\mathsf{H}}

\def\sfP{\mathsf{P}}

\def\sfR{\mathsf{R}}
\def\sfS{\mathsf{S}}
\def\sfT{\mathsf{T}}

\def\sfV{\mathsf{V}}

\def\sfZ{\mathsf{Z}}
\def\bydef{:=}

\def\sfj{{\mathsf{j}}}

\def\sfm{{\mathsf{m}}}

\def\sf0{{\mathsf{0}}}

\def\rmT{\mathrm{T}}

\def\bydef{:=}

\def\rmc{{\mathrm{c}}}

\def\rm0{{\mathrm{0}}}

 \def\bphi{{\pmb{\phi}}}
\def\bGamma{{\pmb{\Gamma}}} 
 
\def\bTheta{{\pmb{\Theta}}} 
 \def\bPsi{{\pmb{\Psi}}}
 
\def\b0{{\pmb{0}}} 
\def\NT{{N_\mathrm{T}}}
\def\NR{{N_\mathrm{R}}}
\def\NS{{N_\mathrm{S}}}

\def\vec{\mathrm{vec}~}

\def\bFrak{{\bF_{\mathsf{ra},k}}}
\def\bFanak{{\bF_{\mathsf{ana},k}}}
\def\bFdigk{{\bF_{\mathsf{dig},k}}}

\newcommand{\trace}[1]{\textrm{Tr}\left({#1}\right)}
\newcommand{\norm}[1]{\left\|{#1}\right\|}
\newcommand{\diagg}{\mathrm{diag}}

\newcommand{\red}[1]{{\color{red}{#1}}} 

\begin{document}
	
	
	
	\title{Signal Processing Foundations of Reconfigurable Antennas in the Tri-Hybrid MIMO Architecture
	}
	
	\author{\IEEEauthorblockN{Nitish Vikas Deshpande, \textit{Graduate Student Member, IEEE,} Joseph Carlson, \textit{Graduate Student Member, IEEE,} Siyun Yang,~\textit{Graduate Student Member, IEEE,} Mohamed Akrout, \textit{Member, IEEE,} Alfredo Gonzalez, \textit{Graduate Student Member, IEEE,}  Miguel Rodrigo Castellanos, \textit{Member, IEEE,} Tharmalingam Ratnarajah,~\textit{Senior Member, IEEE,} Chan-Byoung Chae*,~\textit{Fellow, IEEE}, and Robert W. Heath Jr.*, \textit{Fellow, IEEE}}\\ \vspace{1.5em} \textit{Invited Paper}
		\thanks{ {N.  V. Deshpande, J. Carlson,  and R. W. Heath Jr. are
				with the Department of Electrical and Computer Engineering,  University of California San Diego, La Jolla, CA 92093 USA (email: \{nideshpande, j4carlson,  rwheathjr\}@ucsd.edu). S. Yang and C.-B. Chae are with the School of Integrated Technology, Yonsei University, Seoul 03722, South Korea (email: \{siyun.yang, cbchae \}@yonsei.ac.kr. } A. Gonzalez is with the Joint Doctoral Program in Electrical and Computer Engineering between San Diego State University, San Diego, CA 92182 USA, and the University of California San Diego,
				La Jolla, CA 92093 USA (email: alg113@ucsd.edu). T. Ratnarajah is with the Department of Electrical and Computer Engineering, San Diego State University, San Diego, CA 92182 USA (email: T.Ratnarajah@ieee.org).
				Mohamed Akrout and  Miguel Rodrigo Castellanos are with the Min H. Kao Department of Electrical Engineering and Computer Science, University of Tennessee, Knoxville, TN 37996 USA (e-mail: {makrout, mrcastellanos}@utk.edu).
			This material is also based upon work supported by the National Science Foundation under grant nos. NSF-ECCS-2435261, NSF-CCF-2435254, NSF-ECCS-2414678, and the Army Research Office under Grant W911NF2410107. This work is  also supported in part by NRF and IITP grants funded by Korea Government (RS-2024-00428780, RS-2026-25489110). 
		}
	}

	\maketitle
	
	\begin{abstract}
		To enable larger apertures in multiple-input multiple-output (MIMO) systems, the tri-hybrid MIMO architecture offers a promising low-cost and low-power solution by introducing reconfigurable antennas as a third layer of precoding on top of conventional digital and analog processing. In this paper, we develop a unified signal processing framework for tri-hybrid MIMO that explicitly captures the electromagnetic (EM) characteristics of diverse reconfigurable antenna technologies. We first propose a generic input–output model that incorporates the reconfigurable antenna layer into an effective channel representation, revealing a fundamental coupling between the channel, precoder, and radiated power. Building on this model, we formulate a general optimization problem that jointly accounts for digital, analog, and antenna-domain precoding under hardware and power constraints. We then instantiate this framework across seven representative reconfigurable antenna architectures, including parasitic arrays, dynamic metasurface antennas, fluid/pixel antennas, polarization-reconfigurable antennas, stacked intelligent metasurfaces, pinching antenna systems, and non-radiating wires. To systematically compare these heterogeneous architectures, we introduce a new metric, the reconfigurability efficiency factor (REF), which quantifies the performance gains achievable through antenna reconfiguration under realistic constraints. Numerical results demonstrate the trade-offs among aperture size, power consumption, hardware complexity, and spectral efficiency. Our results establish that EM-level reconfiguration reshapes the signal processing design space, highlighting the need for new architectures and algorithms that jointly optimize across digital, analog, and electromagnetic domains. This work reveals that electromagnetic reconfiguration couples the channel and precoder, transforming the signal processing paradigm for large-aperture MIMO systems.

	\end{abstract}
	
	
	\section{Introduction}\label{sec: Introduction }

	
	\IEEEPARstart{M}{ultiple}-input-multiple-output (MIMO) is the key technology in modern wireless communications, that enables high data rates through spatial multiplexing. While its origin lies in early progress on adaptive arrays, smart antennas, and space-time coding, MIMO evolved into a primary research field by the late 1990s\cite{Paulraj1994}. This evolution sparked diverse and sophisticated subfields, ranging from multi-user MIMO, which facilitates simultaneous communication with multiple devices, to massive MIMO, which employs large-scale antenna arrays to achieve higher capacity and reliability. Today, MIMO is an indispensable component of the global connectivity landscape, integrated into millimeter-wave systems and forming the backbone of high-performance wireless local area network (WLAN) and cellular standards like 5G and beyond.
	
	This paper studies tri-hybrid MIMO, which extends conventional hybrid MIMO by adding a third precoding layer implemented in the antenna domain\cite{11270996}. A standard hybrid architecture combines a digital baseband precoder and an analog RF precoding network; tri-hybrid MIMO places reconfigurable antennas as an additional precoding stage after that analog network. The reconfigurable antenna is summarized by a precoding matrix that alters the effective MIMO channel seen by the baseband, while electromagnetic tuning and mutual coupling also impact the total radiated power, so the channel model and the power constraint are coupled to the same tuning parameters. The architecture therefore unifies the signal processing view of MIMO with the hardware and electromagnetics view of reconfigurable apertures, at the cost of architecture-specific constraints that are absent from a conventional hybrid transceiver.
	
	Section~\ref{subsec:mimo_arch_evolution} reviews MIMO as a line of work on capacity, multi-user operation, and hybrid precoding. Section~\ref{subsec:reconfig_ant_evolution} reviews reconfigurable antennas in the antenna and electromagnetics literature, including parasitic structures, metasurfaces, and fluid- and pixel-based realizations. Section~\ref{subsec:third_precoding_layer} formulates the convergence point: a  tri-hybrid precoding architecture with shared system-level objectives and antenna-specific electromagnetic constraints.

	\subsection{Evolution of MIMO architectures}\label{subsec:mimo_arch_evolution}
	
	By deploying multiple antennas at both ends of a wireless link, a transmitter can spatially multiplex several co‑channel signals and a receiver can separate them using low‑complexity Bell Laboratories Layered Space-Time (BLAST)-type architectures~\cite{Foschini1996}\cite{Wolniansky1998}. This spatial multiplexing significantly enhances spectral efficiency, which scales with the minimum number of antennas available at either end of the link \cite{Foschini1998},\cite{Telatar1999}.
	
	This point‑to‑point MIMO communication model extends to multi‑user MIMO (MU‑MIMO), where the receive‑side antennas are effectively distributed across multiple single‑antenna users to support downlink transmission. Intra‑cell interference can be mitigated by applying precoding at the base station, allowing a multi‑antenna transmitter to serve multiple terminals on the same time‑frequency resource. A corresponding multi‑user uplink configuration is achieved through linear or nonlinear multi‑user detection at the base station, enabling efficient reception of signals from multiple users \cite{Vishwanath2003}.
	
	The 3GPP Release 11 (4G LTE‑A standard) introduced the concept of Coordinated Multi‑Point (CoMP) transmission to enable multi‑cell MU‑MIMO operation under universal frequency reuse. In this framework, base‑station cooperation can mitigate inter‑cell interference caused by simultaneous co‑channel transmissions in neighboring cells, though this requires high‑bandwidth backhaul links between the cooperating sites. Massive MIMO represents the latest evolution of multi‑cell MU‑MIMO systems, in which base stations are equipped with tens or even hundreds of antennas to form highly directional beams toward intended users. This beamforming capability drastically reduces both intra‑cell and inter‑cell interference without requiring coordination among neighboring base stations~\cite{5595728}. Practical fully digital massive MIMO systems, where each antenna element is equipped with its own dedicated RF chain, typically rely on low‑complexity linear processing techniques such as maximum ratio combining (MRC) and maximal ratio transmission (MRT). While more advanced precoders and combiners, such as zero‑forcing (ZF) and minimum mean‑square error (MMSE), offer superior performance, their computational and hardware complexity make them difficult to implement at large array scales. To overcome this limitation, researchers introduced hybrid precoding as a more hardware‑ and energy‑efficient solution, using a reduced number of RF chains connected to subarrays via phase shifters, a structure particularly advantageous at mmWave frequencies \cite{Alkhateeb2014}.

	
	

	\subsection{Evolution of reconfigurable antennas}\label{subsec:reconfig_ant_evolution}

	Researchers in the antenna community have extensively studied the fundamental modeling, analysis, simulation, design, and proof-of-concept development of different types of reconfigurable antennas.
	Some of the early work on reconfigurable antennas was driven by  military applications like missiles, aircrafts, and radars to provide frequency agility and polarization diversity\cite{schaubert1983frequency}. These initial efforts were based on switching varactor diodes to dynamically change the current distribution on the antenna and modify the electromagnetic properties without requiring an external analog network. 
	Hardware advances such as PIN diodes\cite{PokornyEtAlReconfigurableAntennaArrayTestbed2024}, RF MEMS\cite{ErdilEtAlFrequencyTunableMicrostripPatch2007}, and liquid metal\cite{SoEtAlReversiblyDeformableMechanicallyTunable2009} have enabled many forms of reconfigurable antennas. Recent designs include parasitic arrays, pixel-layer antennas, dynamic metasurface antennas, fluid antennas, movable antennas, stacked intelligent metasurfaces, pinching antennas, meta-lines, and polarization-reconfigurable antennas. We review the evolution of these antenna types and discuss the state of the art.
	
	The concept of parasitic antenna arrays originated from the idea of using passive antenna elements to shape the radiation pattern of a single actively driven antenna without requiring multiple RF chains. The field began with classical Yagi-Uda antennas~\cite{yagi1926projector}, which use directors and reflectors to shape the radiation pattern. Researchers later formalized the theory through reactively controlled directive arrays~\cite{1141852}, now commonly called electronically steerable passive array radiators (ESPARs)~\cite{858918}.
	Another related concept is load-modulated arrays or direct antenna modulation\cite{4753998}, which involves switching the impedance values as function of the transmitted information symbol.  
	A circuit theory approach was used in \cite{6877828} for describing the signal model for a parasitic array with a single RF chain.
	That model was generalized to the multi-active multi-passive antenna arrays in \cite{9200503}.  The parasitic arrays have been used recently in wireless systems to develop low-power and low-cost beamforming architectures~\cite{11241086, 11196009, 11443934}.
	

	Dynamic metasurface antennas are a type of leaky-wave based reconfigurable antenna. Varactors or PIN diodes can tune the radiation characteristics of each slot within the waveguide, enabling beamforming across many slot elements. Researchers initially introduced DMAs by showing how metasurface topologies, like complimentary electric-LC resonators, can function as electrically small antennas. They first established the core physics that relates each DMA element to an effective magnetic dipole in terms of its radiation characteristics and frequency response \cite{pulido2017polarizability}. Researchers soon integrated tunable components into the slot elements to enable dynamic beamforming capabilities \cite{SleasmanEtAlWaveguideFedTunableMetamaterialElement2016}, which led to the development of antenna weight configuration methods \cite{smith2017analysis,carlson2024hierarchical} and experimental prototypes to verify the beamsteering performance \cite{boyarsky2021electronicallya}. Recently, DMAs have gained significant traction in the wireless communications community as a means to build large antenna arrays with low power consumption, where signal processing algorithms have been developed to overcome weight limitations for the DMA elements \cite{WangEtAlDynamicMetasurfaceAntennasMIMOOFDM2021,YouEtAlEnergyEfficiencyMaximizationMassive2022}.
	

	Pinching antenna systems (PASS) is another reconfigurable antenna based on the leaky-wave concept \cite{ding2025flexible,liu2026pinching}. PASS is different from DMAs since they rely on the physical movement and activation of multiple dielectric particles along the waveguide often called pinching antennas \cite{wang2025modeling}, offering a much larger spatial aperture for reconfiguration, unlike DMAs which consist of fixed waveguides with sub-wavelength metamaterial elements whose radiation is controlled electronically via tunable components. Because pinching antennas (PAs) can be dynamically repositioned along this extensive aperture, the system can establish direct short-range line-of-sight links in close proximity to users, especially after using its pinching beamforming \cite{sun2025multiuser} and interference cancellation \cite{sun2025pinching} capabilities.

	
	The fluid antenna concept first appeared in the wireless communications literature in the form of the fluid antenna system (FAS), in which a position-flexible radiator can move or switch among multiple candidate ports within a compact space to exploit small-scale spatial channel variations for diversity and multiplexing gains \cite{Wong2021FAS,Wong2020FASLimits}. This framework established the basic communication-theoretic model for position reconfiguration and demonstrated that substantial performance gains can be achieved even with a single active RF chain \cite{Wong2021FAS,Wong2020FASLimits}. It was later extended to multiuser communications through fluid antenna multiple access (FAMA), where spatial reconfigurability is used to mitigate interference and support massive connectivity with low signal-processing complexity \cite{Wong2022FAMA,Hong2026FAMAReview}. More recently, FAS has evolved into a broader hardware-agnostic reconfigurable antenna paradigm that encompasses liquid-based, movable, and pixel/metasurface-inspired realizations, while also enabling new sparse fluid antenna architectures that expand the virtual aperture and spatial degrees of freedom for sensing and array-processing applications such as direction-of-arrival estimation \cite{New2025FASTutorial,Xu2026SparseFADoA,Hong2026FAMAReview}.
	
	Polarization reconfigurable antennas provide systems with polarization diversity without the use of multiple antennas. Early work on MIMO demonstrated that the use of co-located antennas with orthogonal polarizations could significantly improve system performance in terms of error rates and capacity \cite{NabarEtAlPerformanceMultiantennaSignalingTechniques2002, DongEtAlSimulationMimoChannelCapacity2005}. Polarization reconfigurable antennas offer a space- and cost-efficient alternative by replacing dual-polarized array elements with single-port tunable antennas \cite{GaoEtAlPolarizationAgileAntennas2006}. Past designs used switching components, such as PIN diodes and MEMS, to shift between polarizations by changing the feed position, the element shape or the excitation mechanism \cite{SungEtAlReconfigurableMicrostripAntennaSwitchable2004, YangRahmat-SamiiReconfigurablePatchAntennaUsing2002}. In all cases, switches would lead to different current distributions along the antenna, which would change the polarization. Modern designs have built on these approaches by using novel layouts, more advanced switches, and metamaterials to create reconfigurable antennas with higher numbers of polarization states \cite{LinWongWidebandCircularPolarizationReconfigurable2015,SanoHigakiLinearlyPolarizedPatchAntenna2019,WangEtAlTurnstilePolarizer2025,LiEtAlGrapheneBasedCurrentVector2025}. Polarization reconfigurable arrays offer similar benefits to dual-polarized antennas and may be key in enabling large polarization-diverse MIMO arrays with low power consumption.
	
	Stacked intelligent metasurface
	(SIM) is a lens-like hardware component positioned in
	front of a radiating aperture~\cite{intro_stacked_1}. It consists of multiple metasurface layers
	embedded with ``meta-atoms'' that precisely tune the phase and amplitude of
	local electromagnetic (EM) waves. Modifying the EM wave in this way allows the
	SIM to realize digital precoding approach in the wave domain
	\cite{an2024stacked_transceiver,intro_stacked_1}.
	From a signal processing perspective, each layer applies an element-wise
	complex weight to the propagating wavefield, and the cascade of layers
	realizes a trainable multi-stage linear transform in the analog domain.
	Recent work has
	demonstrated the capabilities of this architecture and studied the
	optimization of achievable rates including our prior work \cite{nassirpour2025sumrate,Papazafeiropoulos2024} and others \cite{bahingayi2025scaling}. Additionally, the SIM is highly cost-effective; by
	processing signals in the wave domain, it reduces the computational load and
	hardware requirements of the digital backend.
	
	Beyond single-mode reconfiguration, several designs have achieved multiple forms of reconfigurability within a single antenna. Tunable, loaded stubs, for example, have enabled joint polarization and frequency reconfigurability \cite{Nguyen-TrongEtAlFrequencyPolarizationReconfigurableStubLoadedMicrostrip2015} as well as joint frequency and pattern reconfigurability \cite{ZainarryEtAlFrequencyPatternReconfigurableTwoElementArray2018}. Reconfigurable transmitarrays have also been designed to support pattern and polarization reconfigurability by actively converting the polarization of incoming waves from one state to another \cite{RanaEtAlDigitallyReconfigurableTransmitarrayBeamSteering2021, HuangEtAlUsingReconfigurableTransmitarrayAchieve2015}. By adding a tunable pixel layer to a single patch element, the design in \cite{RodrigoEtAlFrequencyRadiationPatternPolarization2014} achieved simultaneous frequency, pattern, and polarization reconfigurability. Collectively, these works in \cite{Nguyen-TrongEtAlFrequencyPolarizationReconfigurableStubLoadedMicrostrip2015,ZainarryEtAlFrequencyPatternReconfigurableTwoElementArray2018,RanaEtAlDigitallyReconfigurableTransmitarrayBeamSteering2021,HuangEtAlUsingReconfigurableTransmitarrayAchieve2015,RodrigoEtAlFrequencyRadiationPatternPolarization2014} push the boundaries of reconfigurable antenna research by demonstrating that frequency, pattern, and polarization can all be controlled with a single aperture. A limitation of this body of work, however, is that each design is uniquely tailored to a specific application, employing distinct antenna topologies, hardware, and reconfiguration techniques. This diversity in implementation makes it difficult to develop broader signal processing models that generalize across multi-reconfigurable antenna designs.

	
	
	\subsection{The third-precoding layer: convergence of reconfigurable antennas and MIMO}\label{subsec:third_precoding_layer}
	
	Beyond the challenge of generalizing across antenna designs, reconfigurable antennas offer several features well suited to MIMO systems, though most prior work has evolved independently of the MIMO community. In particular, they can introduce additional electromagnetic degrees of freedom while reducing the number of power-hungry active RF components. They have the potential to replace the power hungry fully-digital hardware architecture, similar to how analog phase-shifters were used in the hybrid architecture in 5G NR to offload the precoding complexity. We envision the convergence of these advances in reconfigurable antennas and the sophisticated precoding algorithms developed for MIMO systems in the form of the tri-hybrid MIMO architecture\cite{11270996}. This convergence is also timely because the push toward upper-midband cellular deployments motivates larger base-station apertures while operators seek to reuse existing site density and upgrade installed infrastructure rather than densify the network further. The reconfigurable antennas can be integrated into the MIMO system model and design as a third layer in the precoding architecture.
	
	Integrating reconfigurable antennas into MIMO systems opens new research challenges. In the tri-hybrid MIMO architecture, the radiated power modeling becomes non-trivial because of the tight dependence on the digital precoding, analog precoding, and this third layer of reconfigurable antenna precoding.  The modeling and optimization challenges are different compared to traditional hybrid precoding because of the complicated hardware constraints arising from the reconfigurable antenna precoding. While prior work on reconfigurable antennas focus on optimizing metrics like directivity or radiation efficiency, there is a need to develop new optimization techniques to configure these antennas using system-level metrics like achievable rate or energy efficiency.

	
	The tri-hybrid MIMO architecture appeared as a generalization of the hybrid beamforming architecture with an emphasis on dynamic metasurface antennas in \cite{10476911}\cite{11203239}.   The tri-hybrid MIMO architecture in \cite{liu2025tri} used an analog network of switches and low-resolution phase shifters with the EM precoder modeled as a selection vector across different radiation patterns.
	That work proposed a hierarchical design where the analog, digital, and electromagnetic (EM) precoders are configured at different timescales.
	Another example includes a radiation-center reconfigurable array implemented using reconfigurable pixel antennas from \cite{li2025tri}.
	A multi-user system model was proposed in \cite{zheng2025tri} that integrates pattern reconfigurable antennas in the tri-hybrid framework.
	Fully-digital and hybrid analog-digital architectures allow optimization algorithms to be standardized across antenna types. A tri-hybrid MIMO design, by contrast, must incorporate the unique constraints of each reconfigurable antenna type and develop specialized models and algorithms for reconfiguration. 
	
	The tri-hybrid MIMO formulation provides a unified framework for investigating different reconfigurable antenna types. It enables a systematic comparison of their similarities, differences, and tradeoffs within a MIMO system.
	To summarize, our prior work in \cite{11270996} established the idea of the tri-hybrid MIMO architecture, which has been used for developing new theory for specific reconfigurable antenna types in \cite{10476911, 11203239, liu2025tri, li2025tri, zheng2025tri }. This paper addresses the key research directions outlined in \cite{11270996}. These directions include accurate and unified modeling of diverse reconfigurable antenna types, physically-consistent power and radiation modeling across digital, analog, and antenna layers, and tradeoff analysis frameworks across multiple system-level metrics.
	Concurrent survey-style papers on reconfigurable antennas have also appeared in \cite{zheng2025reconfigurable6g,liu2025reconfigurable}. These works provide broad overviews of reconfigurable-antenna technologies, prototypes, architectures, and signal-processing challenges. By contrast, our objective here is not a comprehensive technology survey, but a unified transceiver-level formulation in which diverse reconfigurable antennas are modeled as a third precoding layer within MIMO, together with architecture-specific effective-channel and radiated-power models and a common tradeoff-analysis framework based on the REF.
	
	Reconfigurable intelligent surfaces (RIS) provide a related but distinct way to introduce electromagnetic programmability into MIMO systems\cite{8910627}. RIS is typically deployed as a separate surface in the propagation channel, whereas tri-hybrid MIMO integrates the reconfigurable layer into the transceiver as an additional precoding stage. Beyond-diagonal RIS (BD-RIS)~\cite{11311540} and holographic/CAPA MIMO~\cite{10232975,deng2023reconfigurable} are also closely related, but they usually rely on surface-based or continuous-aperture abstractions, while this paper focuses on discrete transceiver-integrated antenna-domain precoding with joint coupling across the digital, analog, and electromagnetic layers. Except where an architecture is inherently near-field-oriented, the unified models in this paper adopt the conventional far-field MIMO abstraction; extending the framework to spherical-wave near-field and beam-focusing regimes remains an important direction for extremely large and continuous apertures~\cite{cui_channel_2022-1,ZhangEtAlBeamFocusingNearFieldMultiuser2022}. Unlike conventional hybrid MIMO, the tri-hybrid architecture introduces a fundamental coupling between the electromagnetic configuration and the effective channel, requiring a rethinking of signal processing design from first principles.
	
	\subsection{Contributions and organization}
	
	
	
	We first position this paper relative to prior tri-hybrid MIMO work. Our prior work in \cite{10476911, 11203239} introduced the tri-hybrid idea with a focus on dynamic metasurface antennas and developed the associated beamforming and precoding. The tri-hybrid formulation in \cite{liu2025tri} replaces the conventional analog stage with a switched low-resolution phase-shifter network and represents the electromagnetic precoder as a selection across radiation patterns, while \cite{li2025tri} develops a radiation-center reconfigurable pixel array and \cite{zheng2025tri} extends the framework to multi-user pattern-reconfigurable arrays. Each of these works targets a specific antenna class and its associated algorithms. By contrast, this paper takes a unified transceiver-level viewpoint for the tri-hybrid MIMO architecture, in line with the research directions laid out in our prior work \cite{11270996}. We describe diverse reconfigurable antennas as a common third precoding layer and build the effective channel and radiated power models that are needed for a  comparison across technologies.
	
	We propose a generic transmit-side input-output model for the tri-hybrid MIMO architecture in which the reconfigurable antenna layer appears as an explicit precoding matrix that maps into an effective MIMO channel. On top of this model, we formulate a tri-hybrid MIMO optimization problem over the digital, analog, and reconfigurable antenna precoders that jointly captures mutual information and total radiated power under hardware constraints. We show that the resulting coupling between the effective channel and the radiated power is a structural feature of tri-hybrid MIMO.  We then specialize this framework to seven antenna types, namely, parasitic arrays, pixel and fluid antenna systems, dynamic metasurface antennas, polarization-reconfigurable antennas, stacked intelligent metasurfaces, pinching antenna systems, and non-radiating wires. For each type, we provide the effective-channel map and the radiated-power expression required by the optimization formulation and discuss how the associated hardware constraints shape the feasible set. This specialization makes it clear that the reconfigurable antenna layer is not interchangeable with an additional analog precoding network, since each antenna class imposes its own feasibility and power structure on the tri-hybrid design.
	
	To quantify the benefit of reconfiguration itself, we propose a new metric called the reconfigurability efficiency factor (REF) that compares a reconfigurable tri-hybrid design to a matched non-reconfigurable benchmark under consistent power and hardware accounting. We use the REF to compare different designs of the same antenna type on a common scale after normalizing for baseline aperture and power. The numerical REF examples across antenna families in Section~\ref{sec: numerical examples ref} use different benefit--cost pairings by design; they are not intended to support ordering antenna classes by REF magnitude alone. We illustrate the metric numerically for parasitic arrays, dynamic metasurface antennas, stacked intelligent metasurfaces, and polarization-reconfigurable arrays, which shows that the tradeoffs among spectral efficiency, power consumption, and hardware complexity depend on which electromagnetic constraints dominate in each architecture.
	
	We discuss the organization of this paper. Section~\ref{sec: the tri-hybrid mimo architecture} introduces the tri-hybrid MIMO architecture and presents the general input-output system model. Section~\ref{sec: reconfig antennas} describes the reconfigurable antenna types that can be integrated into this framework, together with their signal models, power constraints, and tradeoffs. Section~\ref{sec: Efficiency of reconfigurability: A new figure of merit for the tri-hybrid MIMO architecture} introduces the reconfigurability efficiency factor and discusses its interpretation. Section~\ref{sec: numerical examples ref} provides numerical REF examples for parasitic arrays, DMAs, SIMs, and polarization reconfigurable antennas. Finally, Section~\ref{sec: key takeaways} summarizes the key takeaways of the paper.

	\textit{Notation}:  A bold lowercase letter $\bz$ denotes a column vector, 	a bold uppercase letter $\bZ$ denotes a matrix, $|\cdot|$  indicates absolute value, $\angle(z)$ denotes argument of a complex number~$z$, $\odot$ denotes the Hadamard (element-wise) product, $\mathrm{blkdiag}(\bA_1,\ldots,\bA_n)$ denotes the block-diagonal matrix with blocks $\bA_1,\ldots,\bA_n$, $(\cdot)^\rmT$ denotes transpose,  $(\cdot)^{\ast}$ denotes conjugate transpose,  $(\cdot)^{\rmc}$ denotes conjugate, $\| .\|$ denotes the L2 norm, $\|.\|_\mathsf{F}$ denotes the Frobenius norm,	$\cR\{z\}$ denotes real part of a complex number $z$, $\cI\{z\}$ denotes  imaginary part of a complex number $z$, $ \bI_N$ represents the identity matrix of size $N$,  $\mathcal{N}_\bbC(0,1)$ represents the complex normal zero mean unit variance random variable.
	
	\section{The tri-hybrid MIMO architecture}\label{sec: the tri-hybrid mimo architecture}
\begin{figure*}
		\centering
		\includegraphics[width=0.96\linewidth]{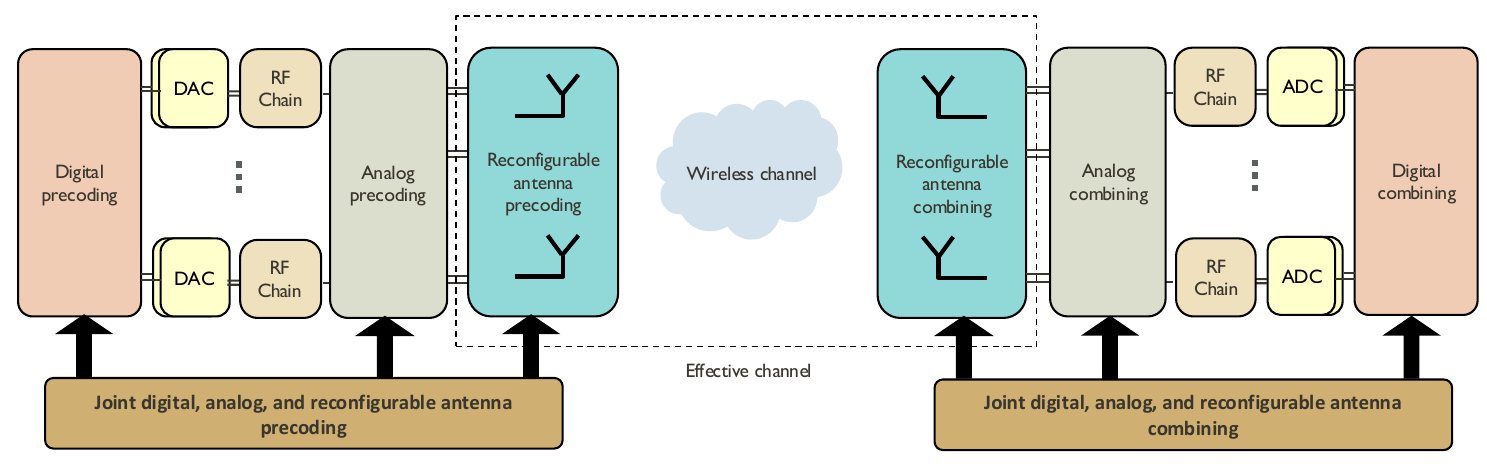}
		\caption{The tri-hybrid MIMO architecture makes use of reconfigurable antennas as a third layer of precoding and combining to augment the capabilities of the legacy hybrid MIMO architecture. The properties of the reconfigurable antenna can be incorporated in an effective channel in the model formulation. The specific system models for each reconfigurable antenna type are shown in their respective subsections in Section~\ref{sec: reconfig antennas}.}
		\label{fig: tri-hybrid mimo}
	\end{figure*}

	The tri-hybrid MIMO architecture generalizes conventional hybrid beamforming by introducing a reconfigurable antenna layer as an additional stage of electromagnetic precoding as shown in Fig.~\ref{fig: tri-hybrid mimo}. This added layer makes it possible to capture diverse reconfigurable antenna technologies within a common signal processing framework while preserving the role of the digital and analog precoders.
	The reconfigurable antenna precoding layer brings in several challenges as it directly impacts the effective wireless channel and the total radiated power through the reconfigurability at the electromagnetic layer.
	We next formalize this architecture through a generic input-output model that serves as the basis for the antenna-specific formulations in Section~\ref{sec: reconfig antennas}.

	
	

	\subsection{Generic input-output system model for tri-hybrid MIMO}\label{subsec: generic io tri-hybrid}
	
	In this section, we provide the  input-output signal model for a generic tri-hybrid MIMO orthogonal frequency-division multiplexing (OFDM) system.
	We focus on the description of the tri-hybrid precoding architecture at the transmitter.   For the receiver, we assume an arbitrary combining architecture.  The tri-hybrid concept could be applied to the receiver too which we defer to future work.
	We assume an OFDM system with $K$ subcarriers.  The subscript $k$ is used to denote the $k$th subcarrier.
	Let the information signal vector of dimension $N_\sfS$ be denoted as $\bx_k \sim \mathcal{CN}(\bm{0}, \bI)$. The digital precoding matrix is denoted as $\bFdigk$. The analog precoding matrix is denoted as $\bFanak$.
	Let the precoding matrix for the reconfigurable antenna array be captured in the matrix $\bFrak$. We do not specify the dimensions of $\bFdigk$, $\bFanak$, and $\bFrak$ because there can be several variations in the architecture. For example, the analog precoding could correspond to phase-shifters, switches, true-time-delays, or a combination of these in  the form of fully-connected, sub-connected, or dynamic sub-array form.  The reconfigurable antenna can be chosen from any of the several types discussed in Section~\ref{sec: reconfig antennas} and could also vary in terms of the array dimension, geometry,  and reconfiguration mechanism. For the sake of comparison, we only fix the number of digital streams input to the digital precoder as $N_\sfS$.
	
	The wireless channel model is  dependent on the antenna and array properties. For a conventional array, the MIMO channel dependence can be modeled through array steering vectors, and the antenna element gains at both ends.  For a reconfigurable antenna array, the dependence of the MIMO channel on the antenna properties is tightly coupled and also varies for different antenna types.
	The heterogeneity and variety in the reconfigurable antennas makes it difficult to define a fixed wireless propagation channel matrix for all antenna types. Instead, we fix the underlying parameters for small scale fading like the delay, angles of arrival and departure, power for each path in a geometric multi-path channel and the parameters for the large scale fading like the distance between the transmitter and receiver.  Let these wireless propagation parameters be denoted by the set $\cH$. 
	Let the antenna specific properties like the antenna type, array dimension, and geometry be denoted by the set $\cA$. 
	
	The effective MIMO channel denoted as $\bH_{\mathsf{eff},k}$  depends on $\cH$, $\cA$, and the reconfigurable antenna precoding matrix $\bFrak$. 
	As only $\bFrak$ is dynamically reconfigurable,  we denote the dependence of $\bH_{\mathsf{eff},k}$ on $\bFrak$ as $\bH_{\mathsf{eff},k}(\bFrak) $.
	The dimension of $\bH_{\mathsf{eff},k}$ is also not fixed but we fix the number of antennas on the receiver to be $N_\sfR$.
	Let the noise vector at the receiver be $\bn_k \sim \mathcal{CN}(\bm{0}, \sigma^2 \bI)$.  The received signal denoted as $\by_k$ is expressed as
	\begin{align}\label{eqn:  generic input output system model}
		\by_k = \bH_{\mathsf{eff},k}(\bFrak) \bFanak\bFdigk\bx_k+ \bn_k.
	\end{align}
	The dependence of $\bH_{\mathsf{eff},k}$ on the reconfigurable antenna precoder $\bFdigk$ is in general a non-linear function, which complicates the system model. This dependence also varies with the reconfigurable antenna type. The specific mathematical form for each type is discussed in Section~\ref{sec: reconfig antennas}.
	
	\subsection{Tri-hybrid MIMO optimization problem formulation}
	
	The general approach to configure the tri-hybrid MIMO precoders is to maximize the average of mutual information over $K$ subcarriers expressed as 
	\begin{align}
		&\cI(\bFrak, \bFanak, \bFdigk)  =\frac{1}{K}\sum_{k=1}^K \log_2\bigg(\bigg|\bI+\nonumber  \\&\frac{1}{\sigma^2} \bH_{\mathsf{eff},k}(\bFrak)\bFanak\bFdigk \bF^\ast_{\mathsf{dig},k}\bF^\ast_{\mathsf{ana},k}\bH^\ast_{\mathsf{eff},k}(\bFrak)\bigg| \bigg).
	\end{align}
	The tri-hybrid precoders are subject to several constraints whose mathematical form would vary with the architecture type. In this formulation, we provide generic constraints without specifying the exact mathematical form.  The specific expressions for these constraints are deferred to Section~\ref{sec: reconfig antennas}.
	
	\begin{figure*}[t]
		\centering
		\includegraphics[width=0.98\linewidth]{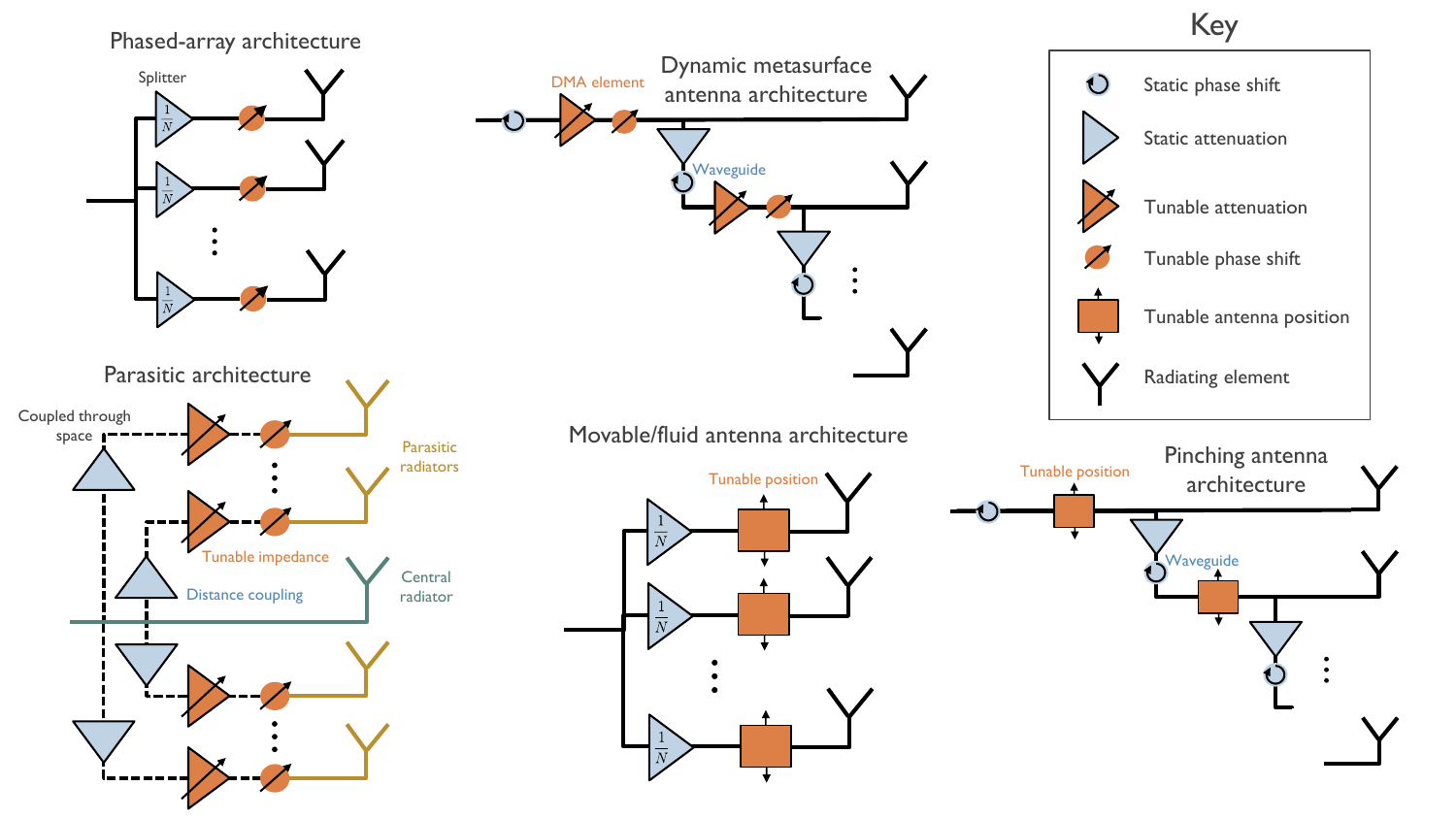}
		\caption{Effective block diagrams for the phased-array baseline and the reconfigurable antenna architectures in Section~\ref{sec: reconfig antennas}. Each architecture is decomposed into tunable and static components for phase, amplitude, and position. The parasitic and DMA architectures share load-based tuning, while pixel arrays and pinching antennas rely on position reconfiguration. These shared abstractions motivate the unified tri-hybrid signal model in Section~\ref{sec: the tri-hybrid mimo architecture}.}
		\label{fig: overview}
	\end{figure*}

	Let the radiated power from the transmitter at the $k$th subcarrier be defined using a general function $P_k(\bFrak, \bFanak, \bFdigk)$. We bound the sum of power across all $K$ subcarriers  by $P_\mathsf{max}$.
	Let the feasible set of analog precoders be $\cF_{\mathsf{ana}}$ and that for the reconfigurable antenna be $\cF_{\mathsf{ra}}$. Using these definitions, we express the optimization problem  to configure the tri-hybrid precoder as
	\begin{subequations}
		\begin{alignat}{3}
			&\underset{\bFrak, \bFanak, \bFdigk}{\mbox{max}} \cI(\bFrak, \bFanak, \bFdigk), \label{eqn: reformualted mutual info obj}\\
			&\text{ s.t. }\sum_{k=1}^K P_k(\bFrak, \bFanak, \bFdigk)\leq P_\mathsf{max}, \label{eqn: tx power constraint}\\
			& \bFanak  \in \cF_{\mathsf{ana}}\label{eqn: analog precoder},\\
			& \bFrak  \in \cF_{\mathsf{ra}}\label{eqn: antenna precoder}.
		\end{alignat}
	\end{subequations}
	This problem is substantially more challenging than conventional hybrid precoding because the reconfigurable antenna layer introduces nonconvex, architecture-dependent constraints that directly affect both the channel and radiated power. In Section~\ref{sec: reconfig antennas}, we discuss specific problem formulations for each type of reconfigurable antenna.

	We elaborate this formulation for a special case where the reconfigurable antenna precoder can be separated from the wireless channel.
	Let $\bH_k$ be the wireless channel that is purely a function of the wireless propagation parameters $\cH$ and the  antenna properties $\cA$. For this special case $ \bH_{\mathsf{eff},k}(\bFrak)= \bH_k \bFrak$, we  configure the tri-hybrid precoders  similar to  the foundational work on hybrid precoding from \cite{ayach_spatially_2014} and \cite{7448873}.
	Let $\bF_{\mathsf{opt},k}$ be the optimal precoder assuming a fully-digital architecture.  The alternate problem formulation is based on minimizing the average of the  Euclidean distance between the optimal precoder $\bF_{\mathsf{opt},k}$ and the actual tri-hybrid precoder $\bFrak \bFanak\bFdigk$ over all subcarriers. Mathematically, we have $\bF^\star_{\mathsf{dig},k}, \bF^\star_{\mathsf{ana},k}, \bF^\star_{\mathsf{ra},k}$
	\begin{subequations}\label{eqn:  alternate formulation}
		\begin{alignat}{3}
			&=\underset{\bFrak, \bFanak, \bFdigk}{\mbox{argmin}} \sum_{k=1}^K\| \bF_{\mathsf{opt},k}- \bFrak \bFanak\bFdigk \|_{\mathsf{F}}^2, \label{eqn: mse trihybrid weight matching}\\
			&\text{ s.t. (3b), (3c), (3d).}  
		\end{alignat}
	\end{subequations}
	Existing solutions cannot be directly applied because of the inclusion of the reconfigurable antenna precoder. As the antenna reconfiguration usually happens through tuning of intrinsic frequency-selective components like variable capacitances, the values of $\bFrak$ are not independent across frequencies unlike $\bFdigk$.
	Solving this optimization problem with tightly coupled constraints  opens up new challenges for the MIMO community.


	\section{Reconfigurable antennas for the tri-hybrid MIMO architecture}\label{sec: reconfig antennas}
	
	In this section, we describe the working mechanism, models, and tradeoffs  for different reconfigurable antenna architectures. 
	We provide block diagrams for different reconfigurable antenna types and the phased-array benchmark in  Fig.~\ref{fig: overview}.
	The phased-array model consists of a tunable phase shift for each radiating element. The reconfigurable antenna types have additional tunable elements like a tunable attenuation or a tunable radiating element position. Usually, the tunable attenuation and tunable phase shift are coupled which limits the beamforming flexibility compared to the phased-array benchmark. The tradeoffs for each reconfigurable antenna type are discussed in their respective subsections. 
	We also connect the model of each antenna type to the generic tri-hybrid MIMO architecture signal model proposed in Section~\ref{subsec: generic io tri-hybrid}.

	\subsection{Parasitic antenna arrays}\label{subsec: parasitic arrays}


	Parasitic antenna arrays are a type of reconfigurable antenna architecture whose working mechanism is based on the mutual coupling phenomenon.  With several densely packed antenna elements in an array, the mutual coupling between neighboring elements is significant~\cite{ivrlac_toward_2010, deshpande2023analysis}. A current on an antenna generates an electromagnetic field which induces a voltage on neighboring elements. If each antenna in a dense array is actively driven by a current source, it can lead to prohibitively higher power consumption.  
	
	To reduce the number of active components in the hardware, an alternative technique is to replace the active current sources on some antenna elements with a tunable load. These elements with a reconfigurable load are termed as ``parasitic" elements as the currents on these elements are induced by actively driven antennas.  This approach of introducing parasitic elements and using their coupling with active elements enables scaling up the number of antennas in MIMO arrays without a significant increase in the number of RF chains.

	We describe the input-output signal model  for a parasitic antenna array based tri-hybrid MIMO.
	We use the parasitic antenna system model developed in our prior work \cite{11241086, 11443934} to provide an expression for  the effective channel term $\bH_{\mathsf{eff},k}(\bFrak)$ in \eqref{eqn:  generic input output system model}.
	The circuit-theoretic approach developed by \cite{ivrlac_toward_2010, deshpande2023analysis, mezghani2023reincorporating, wallace_mutual_2004} is useful for a physically-consistent formulation of the effective channel. 
	With the circuit model, the output signal $\by_k$ physically denotes the voltage at the receiver ports, the effective channel $\bH_{\mathsf{eff},k}(\bFrak)$ has the dimension of impedance, and $\bFanak\bFdigk\bx_k$ represents the current vector of the active antennas.
	
	For a parasitic hybrid array, we assume that there are $N_\sfA$ active and $N_\sfP$ parasitic antennas.
	Let the wireless propagation channel matrix between the receiver and the active antennas be $\bm{\sfZ}_{\mathsf{RA},k} \in \bbC^{N_\sfR \times N_\sfA}$ and that between the receiver and the parasitic antennas be $\bm{\sfZ}_{\mathsf{RP},k} \in \bbC^{N_\sfR \times N_\sfP}$.
	The mutual impedance matrix between all parasitic antennas is $\bm{\sfZ}_{\mathsf{P},k} \in \bbC^{N_\sfP \times N_\sfP}$,  between all active antennas is $\bm{\sfZ}_{\mathsf{A},k} \in \bbC^{N_\sfA \times N_\sfA}$, and that between the parasitic and active antennas is $ \bm{\sfZ}_{\sfm,k} \in \bbC^{N_\sfP \times N_\sfA}$.
	We assume that the $\ell$th parasitic antenna can be tuned with a reconfigurable load $Z_{\mathsf{R},\ell}$.
	The parasitic reconfigurable load matrix is defined as  $\bm{\sfZ}_{\mathsf{R}}=\mathsf{diag}\{[Z_{\mathsf{R},1}, \dots, Z_{\mathsf{R},N_{\mathsf{P}}}]^\rmT\} $. We define the parasitic antenna precoding matrix as $ \bFrak=( \bm{\sfZ}_{\mathsf{P},k} + \bm{\sfZ}_{\mathsf{R}})^{-1}$. The effective MIMO channel in \eqref{eqn:  generic input output system model} is expressed in terms of $ \bFrak$ as
	\begin{equation}\label{eqn: effective MIMO channel parasitics}
		\bH_{\mathsf{eff},k}(\bFrak)  = \bm{\sfZ}_{\mathsf{RA},k} - \bm{\sfZ}_{\mathsf{RP},k}\bFrak\bm{\sfZ}_{\sfm,k} .
	\end{equation}
	For a conventional array without parasitic elements, $\bH_{\mathsf{eff},k}=\bm{\sfZ}_{\mathsf{RA},k}$ would be a fixed matrix.
	The detailed derivation of the effective MIMO channel matrix for the parasitic array is provided in \cite{11241086}.
	
	\begin{figure}
		\centering
		\includegraphics[width=0.5\textwidth]{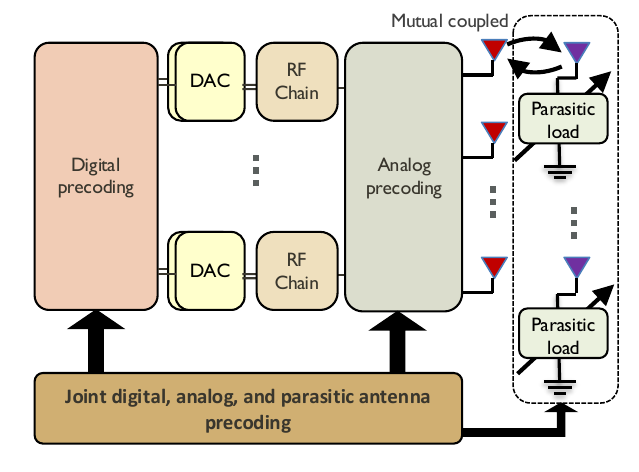}   
		\caption{System model for the parasitic antenna based tri-hybrid architecture. The active (red) and parasitic (purple) antennas are closely spaced, which leads to mutual coupling that augments the beamforming capability beyond the phased-array baseline.}
		\label{fig: parasitic system model}
	\end{figure}

	The circuit-theoretic approach is useful for defining a realistic radiated power constraint. For formulating the power constraint, we define an effective impedance matrix as
	\begin{align}
		&\bm{\sfZ}_{\mathsf{eff},k}=\cR\{\bm{\sfZ}_{\sfA,k} \} -\cR\{ \bm{\sfZ}_{\sfm,k}^\rmT\}\bFrak\bm{\sfZ}_{\sfm,k} \\ \nonumber  &
		-    \bm{\sfZ}_{\sfm,k}^\ast \bFrak^{\ast}  \cR\{\bm{\sfZ}_{\sfm,k}  \}
		+\bm{\sfZ}_{\sfm,k}^\ast \bFrak^{\ast}\cR\{ \bm{\sfZ}_{\sfP,k} \} \bFrak\bm{\sfZ}_{\sfm,k}.
	\end{align}
	Using this definition, the transmitted power at the $k$th subcarrier is expressed as
	\begin{align}
		P_k(\bFrak, \bFanak, \bFdigk) = \mathsf{Tr}(\bF^\ast_{\mathsf{dig},k} \bF^\ast_{\mathsf{ana},k}\bm{\sfZ}_{\mathsf{eff},k}\bFanak\bFdigk).
	\end{align}
	We observe that the transmitted power expression depends on the parasitic precoding matrix $\bFrak$ through $\bm{\sfZ}_{\mathsf{eff},k}$.
	For a conventional hybrid array without any parasitics, and assuming uncoupled active antennas, the power term simplifies to the Frobenius norm squared of the hybrid precoder $\bFanak\bFdigk$~\cite{park_dynamic_2017}.

\begin{figure}[t]
		\centering
		\includegraphics[width=0.48\textwidth]{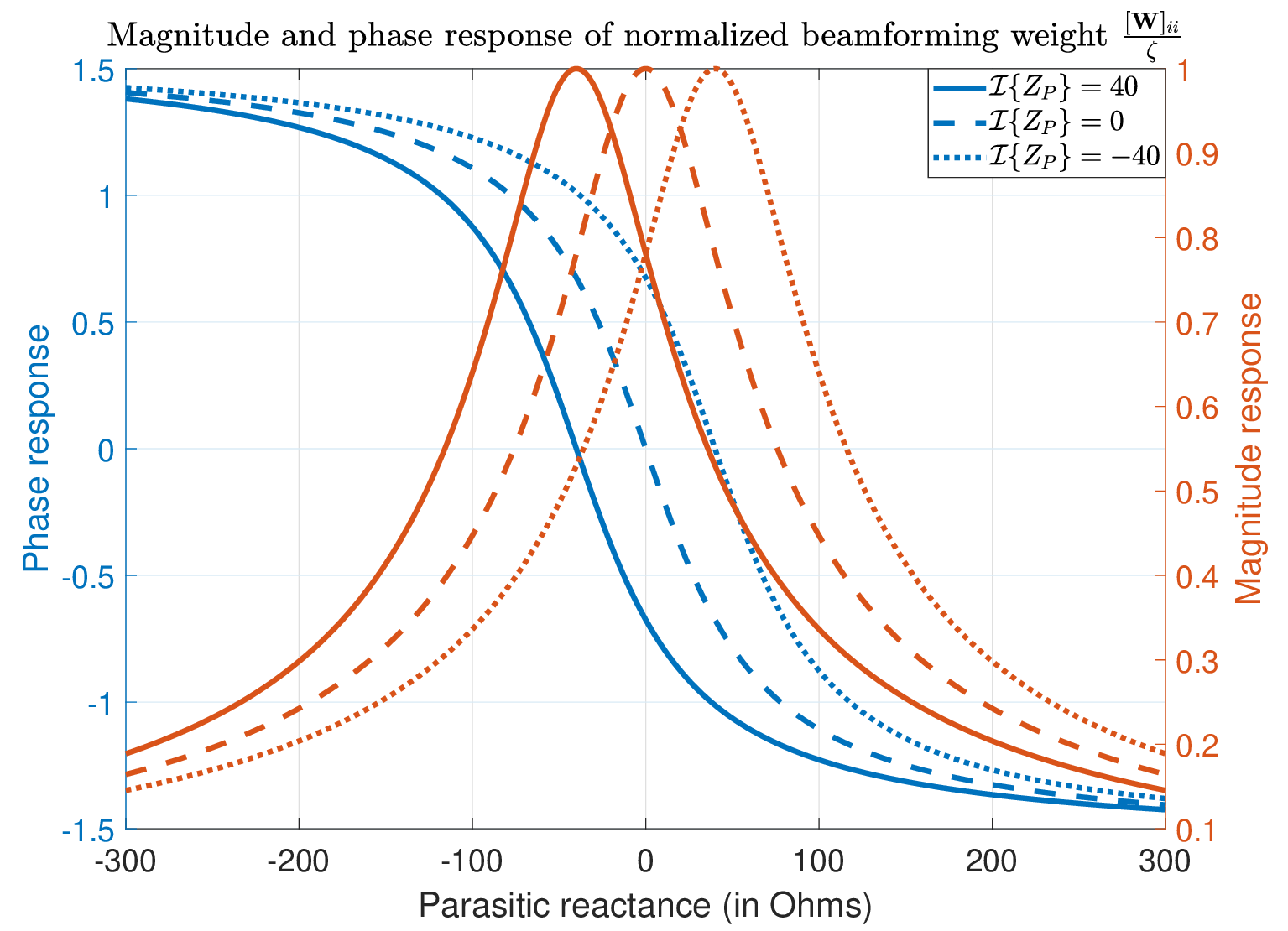}    
		\caption
		{{The normalized beamforming weight magnitude and phase are reconfigurable based on the parasitic reactance. The responses also depend on the self reactance of the parasitic elements which are not reconfigurable but depend only on the antenna design.}
		}
		\label{fig: bf_weight_mag_phase_response}
	\end{figure}

	Introducing parasitic loads on  antennas instead of an active source leads to some tradeoffs.  With an active source, the magnitude and phase of the current can be independently tuned assuming the source is controlled digitally.  With parasitic loads, the control of the current on the parasitic element becomes highly constrained. Specifically, it was shown in \cite{11241086} that there exists a coupled magnitude and phase constraint in the beamforming weights of  parasitic elements.  This limits the beamforming capability compared to a conventional antenna array.
	We provide an interpretation of the beamforming weight of the parasitic antenna array.  For the scalar case, i.e., $N_\sfP=1$ and $N_\sfA=1$, we obtain the scalar $[\bFrak]_{1,1}=\frac{1}{ {\sfZ}_{\mathsf{P},k} + {\sfZ}_{\mathsf{R}}} $. We fix the real part of the parasitic reactance and vary the reactance.  In Fig.~\ref{fig: bf_weight_mag_phase_response}, we plot $[\bFrak]_{1,1} (\cR\{{\sfZ}_{\mathsf{P},k} + {\sfZ}_{\mathsf{R}}  \})$ for different values of $\cI\{ {\sfZ}_{\mathsf{P},k}\}$ and  vary $\cI\{ {\sfZ}_{\mathsf{R}} \}$ on the $x$ axis.  We observe that the curve exhibits a coupled magnitude and phase behavior which we mathematically show in \cite{11241086}. This coupling is also observed in other types of reconfigurable antennas like dynamic metasurface antennas and is termed as the Lorentzian constraint\cite{smith2017analysis,carlson2024hierarchical,deshpande_qif1}. 
	Despite these challenges in beamforming, the savings in the power consumption dominate which lead to a higher energy efficiency compared to conventional antennas~\cite{11241086}.  We discuss the benefits of reconfiguration in Section~\ref{sec: Efficiency of reconfigurability: A new figure of merit for the tri-hybrid MIMO architecture}.

	\subsection{Pixel arrays and Fluid antenna systems}\label{subsec: pixel fas}
	{
		Pixel arrays and fluid antenna systems (FAS) are selection-based reconfigurable antenna architectures. In pixel arrays, a dense grid of conducting pixels is interconnected through a switching network, which creates a finite set of feasible conductive states \cite{song2014ga_pixel,shen2017sebo,jing2022diagonal}. Each state activates a subset of pixels and induces an effective current distribution and radiation pattern. In FAS, the radiating location is reconfigured so that a position-flexible radiator selects one among several candidate ports or locations to exploit spatial sampling and selection diversity \cite{Wong2021FAS,wong2023fas_part1,chai2022port}. Recent work has also connected these ideas through fast-switching pixel realizations of FAS-style position selection \cite{zhang2025prafas} and electronically movable selection-based architectures for multiuser communications \cite{chen2025remaa}. From a signal-processing viewpoint, these architectures are closely related to movable-antenna models that optimize antenna positions directly \cite{zhu2024movable,ma2023mimo_movable}. Here, we adopt a unified model that treats pixel arrays and FAS as discrete state-selection precoders in the tri-hybrid MIMO architecture.
		
		We consider $N_{\ps}$ controllable analog feeds, and each feed has $\Nsub$ candidate states. For the $p$th feed, let $\be_{n_p}\in\{0,1\}^{\Nsub}$ be a one-hot selection vector satisfying $\|\be_{n_p}\|_0=1$. Collecting the per-feed selections yields the block-diagonal state-selection matrix
		\begin{equation}\label{eq: state_selection_matrix}
			\mathbf{S}_{\mathsf{ra}}
			\bydef
			\mathrm{blkdiag}\!\left(\be_{n_0},\be_{n_1},\ldots,\be_{n_{N_{\ps}-1}}\right)
			\in \{0,1\}^{(N_{\ps}\Nsub)\times N_{\ps}}.
		\end{equation}
		The feasible set implied by one-hot state selection is
		\begin{equation}\label{eq: state_selection_feasible_set}
			\begin{aligned}
				\cS_{\mathsf{ra}}
				=
				\big\{
				\mathbf{S}_{\mathsf{ra}} &= \mathrm{blkdiag}(\be_{n_0},\ldots,\be_{n_{N_{\ps}-1}}) \\
				&:\; n_p\in\mathcal{S}_p\subseteq\{0,\ldots,\Nsub-1\},\ \|\be_{n_p}\|_0=1
				\big\}.
			\end{aligned}
		\end{equation}
		The admissible set $\mathcal{S}_p$ captures hardware-feasible states, such as constraints imposed by the switch interconnection graph in pixel arrays \cite{shen2017sebo,jing2022diagonal} or feasible location constraints in FAS \cite{New2025FASTutorial}.
		
		The one-hot vectors in \eqref{eq: state_selection_matrix} select antenna states, not individual pixels. We assume that feed $p$ controls a subpanel consisting of $N_{\mathsf{tx},p}$ physical radiating elements. For the $p$th feed, define the state dictionary
		\begin{equation}\label{eq: per_feed_dictionary}
			\mathbf{D}_p \in \bbC^{N_{\mathsf{tx},p}\times \Nsub},
			\qquad
			\mathbf{D}_p
			=
			\big[\bd_{p,0},\bd_{p,1},\ldots,\bd_{p,\Nsub-1}\big],
		\end{equation}
		where the column $\bd_{p,n}\in\bbC^{N_{\mathsf{tx},p}}$ represents the relative complex excitation pattern across the physical elements in subpanel $p$ induced by state $n$.
		
		The normalization in \eqref{eq: dp_normalization} separates the spatial shape of each activation pattern from the state-dependent loss, thereby avoiding double-counting of insertion loss in the system model. For pixel arrays, a physical switching state naturally defines an on/off activation pattern $\tilde{\bd}_{p,n}\in\{0,1\}^{N_{\mathsf{tx},p}}$, where multiple physical elements may be simultaneously active under a single state. We then normalize this binary activation pattern as
		\begin{equation}\label{eq: dp_normalization}
			\bd_{p,n}
			\bydef
			\frac{\tilde{\bd}_{p,n}}{\|\tilde{\bd}_{p,n}\|_2},
			\qquad
			\|\bd_{p,n}\|_2=1,
			\quad \forall p,n.
		\end{equation}
		Thus, $\bd_{p,n}$ captures only the relative excitation geometry across the subpanel. The state-dependent losses, including impedance mismatch, switch insertion loss, and ohmic dissipation, are modeled separately through the realized-efficiency coefficients introduced later. In a calibrated model, $\bd_{p,n}$ can be complex-valued and capture both phase and relative amplitude variations within the subpanel.
		
		Let
		\begin{equation}\label{eq: ntx_definition_pixel}
			N_{\mathsf{tx}} = \sum_{p=0}^{N_{\ps}-1} N_{\mathsf{tx},p}
		\end{equation}
		denote the total number of physical radiating elements across all subpanels. Stacking the per-feed dictionaries yields the global dictionary
		\begin{equation}\label{eq: global_dictionary}
			\mathbf{D}_{\mathsf{ra}}
			\bydef
			\mathrm{blkdiag}\!\left(\mathbf{D}_0,\mathbf{D}_1,\ldots,\mathbf{D}_{N_{\ps}-1}\right)
			\in \bbC^{N_{\mathsf{tx}}\times (N_{\ps}\Nsub)}.
		\end{equation}
		The reconfigurable antenna precoder that maps the $N_{\ps}$ analog-feed signals to the $N_{\mathsf{tx}}$ physical radiating elements is then
		\begin{equation}\label{eq: fra_from_dictionary_and_selection}
			\mathbf{F}_{\mathsf{ra}}
			\bydef
			\mathbf{D}_{\mathsf{ra}}\,\mathbf{S}_{\mathsf{ra}}
			\in \bbC^{N_{\mathsf{tx}}\times N_{\ps}}.
		\end{equation}
		
		\begin{figure}
			\centering
			\includegraphics[width=0.92\linewidth]{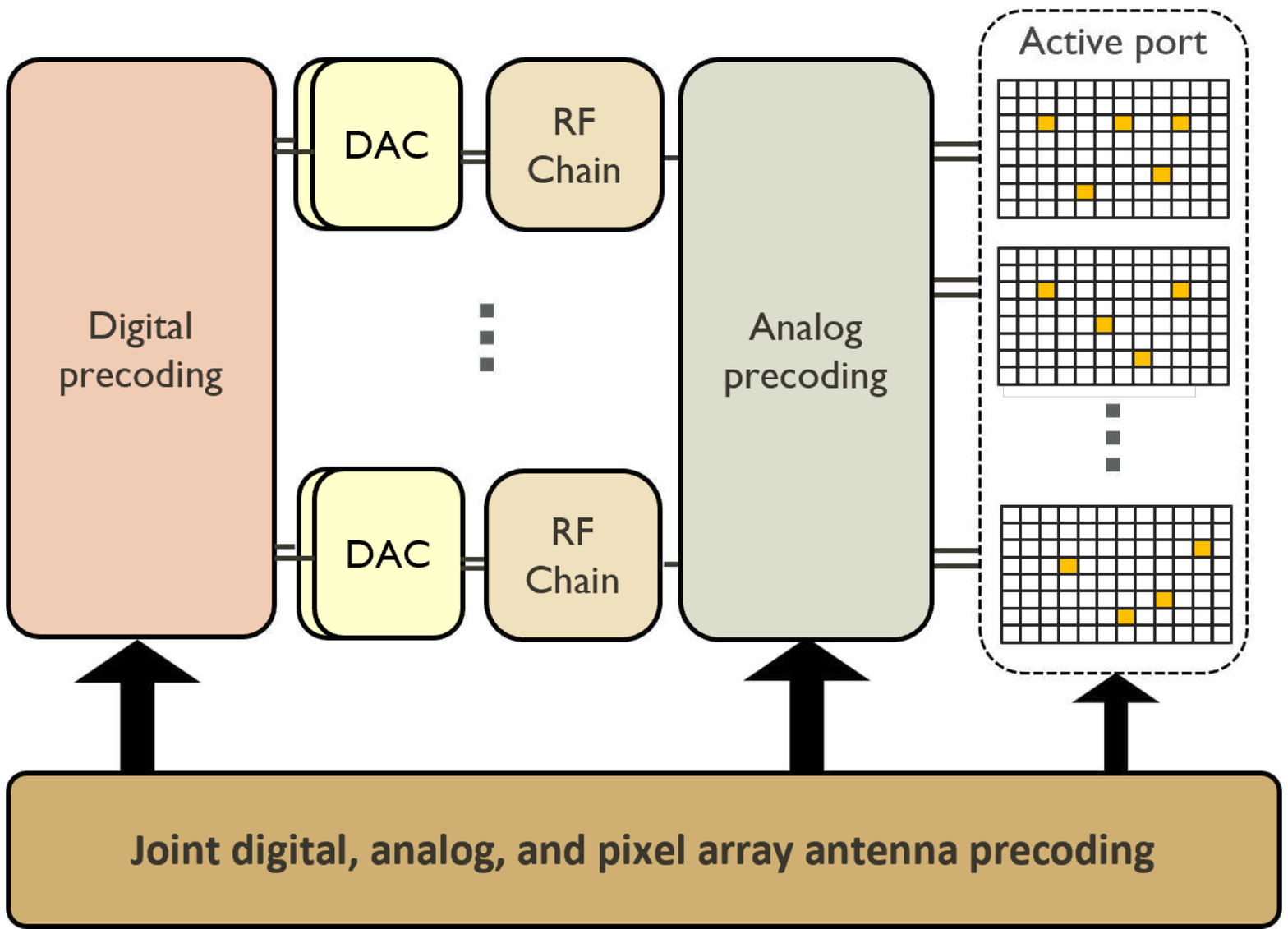}
			\caption{System model for the pixel-array/FAS-based tri-hybrid architecture. Each controllable analog feed selects one state from a finite candidate set through a switching network. The selected state activates a subset of physical radiating elements in the corresponding subpanel and induces the excitation pattern $\bd_{p,n_p}$. }
			\label{fig: pixel system model}
		\end{figure}
		With this definition, the $p$th column of $\mathbf{F}_{\mathsf{ra}}$ equals $\mathbf{D}_p\be_{n_p}=\bd_{p,n_p}$, meaning that a single selected state can activate multiple physical pixels through the pattern $\bd_{p,n_p}$. The overall tri-hybrid precoder is $\mathbf{F}_{\mathsf{ra}}\bFanak\bFdigk$.
		
		Let $\bH_k\in\bbC^{\NR\times N_{\mathsf{tx}}}$ denote the channel from the physical transmit radiating elements to the receiver. Substituting \eqref{eq: fra_from_dictionary_and_selection} into the generic input-output model \eqref{eqn:  generic input output system model} gives the effective channel
		\begin{equation}\label{eq: heff_pixel_fas_physical}
			\bH_{\mathsf{eff},k}(\mathbf{F}_{\mathsf{ra}})
			=
			\bH_k\,\mathbf{F}_{\mathsf{ra}}
			=
			\bH_k\,\mathbf{D}_{\mathsf{ra}}\,\mathbf{S}_{\mathsf{ra}}.
		\end{equation}
		This representation separates discrete state selection through $\mathbf{S}_{\mathsf{ra}}$ from the physical excitation patterns encoded in $\mathbf{D}_{\mathsf{ra}}$.
		
		Selection-based architectures exhibit configuration-dependent impedance mismatch and insertion loss. Since each pattern vector $\bd_{p,n}$ is normalized in \eqref{eq: dp_normalization}, we model these loss effects separately through state-dependent realized efficiencies. Let $\eta_{p,n}\in(0,1]$ denote the realized efficiency associated with state $n$ at feed $p$, and stack these coefficients into the vector
		\begin{equation}\label{eq: eta_stack}
			\begin{aligned}
				\boldsymbol{\eta}
				&=
				\big[
				\eta_{0,0},\ldots,\eta_{0,\Nsub-1},
				\eta_{1,0},\ldots,\eta_{1,\Nsub-1},\\
				&\qquad \ldots,
				\eta_{N_{\ps}-1,0},\ldots,\eta_{N_{\ps}-1,\Nsub-1}
				\big]^\transpose.
			\end{aligned}
		\end{equation}
		
		The equality $\mathbf{F}_{\mathsf{ra}}=\mathbf{D}_{\mathsf{ra}}\mathbf{S}_{\mathsf{ra}}$ implies that each column $[\mathbf{F}_{\mathsf{ra}}]_{:,p}$ of $\mathbf{F}_{\mathsf{ra}}$ is intended to match one of the dictionary columns in $\mathbf{D}_p$. In practice, calibration error or model mismatch can make this relation approximate, so we define the selected state index at feed $p$ as the closest dictionary element:
		\begin{equation}\label{eq: selected_state_index_from_Fra_relaxed}
			n_p(\mathbf{F}_{\mathsf{ra}})
			\in
			\arg\min_{n\in\mathcal{S}_p}
			\big\|[\mathbf{F}_{\mathsf{ra}}]_{:,p}-\bd_{p,n}\big\|_2^2.
		\end{equation}
		If the minimizer is not unique, one of the minimizers is selected arbitrarily.
		
		Using these indices, the per-feed selected efficiency vector is
		\begin{equation}\label{eq: eta_selected_vector_Fra}
			\boldsymbol{\eta}_{\mathsf{sel}}(\mathbf{F}_{\mathsf{ra}})
			\bydef
			\big[
			\eta_{0,n_0(\mathbf{F}_{\mathsf{ra}})},\;
			\eta_{1,n_1(\mathbf{F}_{\mathsf{ra}})},\;
			\ldots,\;
			\eta_{N_{\ps}-1,n_{N_{\ps}-1}(\mathbf{F}_{\mathsf{ra}})}
			\big]^\transpose
			.
		\end{equation}
		We define the corresponding diagonal efficiency matrix as
		\begin{equation}\label{eq: gamma_sel_def_Fra}
			\bGamma(\mathbf{F}_{\mathsf{ra}})
			\bydef
			\mathrm{diag}\!\big(\boldsymbol{\eta}_{\mathsf{sel}}(\mathbf{F}_{\mathsf{ra}})\big)
			\in \bbR^{N_{\ps}\times N_{\ps}}.
		\end{equation}
		
		This allows the radiated-power model to be written directly as a function of the reconfigurable antenna precoder $\mathbf{F}_{\mathsf{ra}}$. A compact model consistent with \eqref{eqn: tx power constraint} is
		\begin{equation}\label{eq: pk_sel_trace_Fra}
			P_k(\mathbf{F}_{\mathsf{ra}},\bFanak,\bFdigk)
			\!=\!
			\mathsf{Tr}\!\left(
			\bFdigk^{\ast}\bF_{\mathsf{ana},k}^{\ast}\,
			\bGamma(\mathbf{F}_{\mathsf{ra}})\,
			\bFanak\bFdigk
			\right)\!.
		\end{equation}
		A scalar-summary efficiency factor can also be defined as
		\begin{equation}\label{eq: eta_ra_avg_new_Fra}
			\eta_{\mathsf{ra}}(\mathbf{F}_{\mathsf{ra}})
			\bydef
			\frac{1}{N_{\ps}}\mathbf{1}^\transpose \boldsymbol{\eta}_{\mathsf{sel}}(\mathbf{F}_{\mathsf{ra}})
			=
			\frac{1}{N_{\ps}}\sum_{p=0}^{N_{\ps}-1}\eta_{p,n_p(\mathbf{F}_{\mathsf{ra}})},
		\end{equation}
		in which case one may use the simplified approximation
		\begin{equation}\label{eq: pk_sel_simplified_new_Fra}
			P_k(\mathbf{F}_{\mathsf{ra}},\bFanak,\bFdigk)
			\approx
			\eta_{\mathsf{ra}}(\mathbf{F}_{\mathsf{ra}})\,\|\bFanak\bFdigk\|_{\Fnorm}^2.
		\end{equation}
		
		Pixel arrays and FAS architectures achieve selection diversity through switching or position reconfiguration. This design trades continuous per-element weight control for a simpler discrete selection mechanism. Their main advantage is reduced RF and analog hardware complexity for a large virtual aperture. Their key drawback is that the selection variable $\mathbf{S}_{\mathsf{ra}}$ is discrete and the radiated power depends on the selected state through $\bGamma(\mathbf{F}_{\mathsf{ra}})$ or $\eta_{\mathsf{ra}}(\mathbf{F}_{\mathsf{ra}})$. This makes the design combinatorial and requires joint consideration of channel strength and state-dependent efficiency.

	}
	
	\subsection{Dynamic metasurface antennas}\label{sec: DMA}
	
	
	Dynamic metasurface antennas are a type of frequency-reconfigurable slotted-waveguide antenna array where tunable slots enable beamforming. By integrating RF components like PIN or varactor diodes into the slots, it is possible to tune the resonant frequency of each slot element independently. Tuning the resonant frequency then creates an adjustable antenna weight at a specific target frequency \cite{smith2017analysis}. DMAs have been shown experimentally to be capable of steering directive main lobes that are similar in gain to a typical antenna array using phase shifters but with much lower power consumption \cite{boyarsky2021electronicallya}.
	
	\begin{figure}
		\centering
		\includegraphics[width=0.45\textwidth]{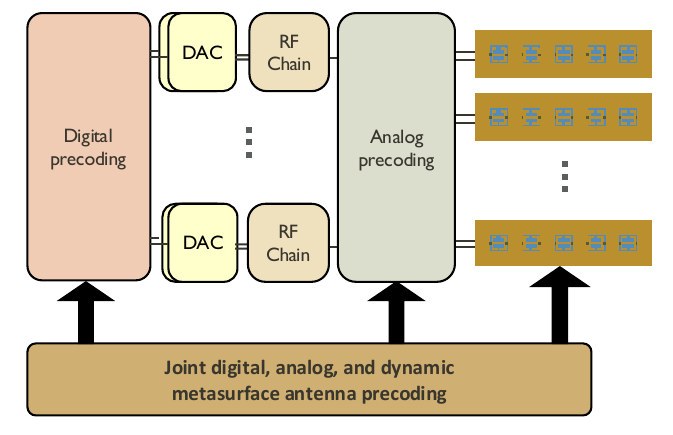}
		\caption{System model for the DMA-based tri-hybrid precoder. Each DMA subarray contains multiple slot elements whose resonant frequencies are independently tuned by varactor diodes. The waveguide-fed structure couples the slot responses through forward scattering, making the effective antenna weights dependent on the configuration of all preceding elements.}
		\label{fig: system model}
	\end{figure}
	
	We now describe the input-output model for the DMA-based tri-hybrid architecture. In the tri-hybrid architecture, we assume several DMA linear subarrays are placed side-by-side to form a full planar array, though other geometric arrangements of the DMAs are also possible. Each DMA subarray connects to a phase shifter, and we assume that $\Pps$ phase shifters are connected to an RF chain. Thus, let each DMA subarray have $\Nsub$ elements that connect to $\Nps$ phase shifters. For the $\ps$th phase shifter, we define the DMA subarray precoding vector as $\bff_{\dma,\ps} = [\rmf_{\dma,\ps\Nsub},\ldots,\mathrm{f}_{\dma,(\ps+1)\Nsub-1}]^\transpose$ and the full DMA precoder as $\bFrak = \text{blkdiag}(\rmf_{\dma,0},\ldots,\rmf_{\dma,\Nps-1})$. Similarly, the $\rf$th precoding vector for each RF chain that consists of $\Pps$ phase shifters is $\bff_{\ana,\rf} = [\rmf_{\ana,\Pps\rf},\ldots,\rmf_{\ana,\Pps\rf+\Pps-1}]^\transpose$ and the full analog precoder is $\bFanak = \text{blkdiag}(\bff_{\ana,0},\ldots, \bff_{\ana,\Nrf-1})$. Lastly, the digital precoder is then $\bFdigk \in \mathbb{C}^{\Nrf \times \NS}$. Since each DMA slot acts as a unique antenna element, the tri-hybrid precoder is linearly separable from the channel such that $ \bH_{\mathsf{eff},k}(\bFrak)= (\bH_k \odot \bG) \bFrak$, where we define $\bG$ later in this section based on the waveguide propagation.

	The effective DMA beamforming weights $\rmf_\dma$ are given by a combination of the antenna weight tunability from the varactor diode, and the forward scattering effects from the electromagnetic fields within the waveguide as each DMA element resonates. As the varactor diode capacitance is tuned, the radiation from the DMA element changes in both amplitude and phase based on its intrinsic magnetic polarizability. A useful abstraction to the magnetic polarizability model is a Lorentzian constraint, which provides a simple and physically-consistent method to model the radiated amplitude and phase as a function of the varactor diode tuning state. Thus, let $\coupling \in (0,1]$ be a term that represents the maximum scattering strength of the DMA elements. For simplicity, we define the $m$th element to have index $m=p\Nsub+n$, where $p \in \{0,\ldots,\Nps-1\}$ and $n \in \{0,\ldots,\Nsub-1\}$. For a phase value $z_m$ that represents the varactor diode tunability, the beamforming weight through the Lorentzian constraint model for each element is then \cite{carlson2025interference}
	\begin{multline}\label{eq: dma beamforming weight}
		\rmf_{\dma,m} = -\frac{\sfj}{2}(1+e^{\sfj z_{m}}) \prod\limits_{\ell=p\Nsub}^{m-1} \left[ 1-\frac{\coupling}{2}(1+e^{\sfj z_\ell})\right], \\ \; z_m\in [-\pi,\pi).
	\end{multline}
	We consider the weights $\rmf_{\dma,m}$ in \eqref{eq: dma beamforming weight} as the reconfigurable antenna precoding vector in the tri-hybrid architecture. 
	
	In addition to the antenna weights, the waveguide also impacts the resulting array response vector of the DMA. Due to the electromagnetic propagation of the fields within the waveguide, there is an inherent phase advance that accumulates at each element position. Thus, let $\btg$ be the waveguide phase propagation constant. We assume the DMA linear subarrays are oriented along the $x$-axis, and the combined subarrays form a planar array along the $y$-axis. The phase advance vector for all DMA elements is then given by 
	\begin{equation}
		\bq = [e^{-\sfj \beta_\mathsf{g} x_{0}}, \ldots, e^{-\sfj \beta_\mathsf{g} x_{\NT-1}}].
	\end{equation}
	To combine with the wireless channel $\bH_k$, we define the phase advance matrix by repeating the phase advance vector $\NR$ times as $\bQ = \textbf{1}_\NR \bq$.
	The effective wireless channel for the DMA-based tri-hybrid architecture is then the combination of the wireless channel and waveguide phase advance as $\bH_k \odot \bQ$.
	
	
	Lastly, we can also model the radiated power based on the forward scattering of the DMA linear arrays. The configurations $z_m$ used to calculate the DMA beamforming weights in \eqref{eq: dma beamforming weight} are also necessary to calculate the forward scattering in the product term. For an input power $\Pink = || \bFdigk \bFanak ||^2_{\Fnorm}$ into each DMA subarray, the DMA elements will radiate most of the power as the electromagnetic fields within the waveguide decay due to the forward scattering. Any leftover power in the waveguide electromagnetic fields after the last DMA element is then absorbed into a matched load and not radiated by the DMA. Therefore, we calculate the radiated power constraint for the DMA by subtracting the leftover power via forward scattering from the initial input power. We define the forward scattered electromagnetic fields after the last DMA element for subarray $\ps$ as
	\begin{equation}
		S_{12,\ps}^{\mathsf{all}} = \prod\limits_{\ell=p\Nsub}^{\Nsub(p+1)-1} \left[ 1-\frac{\coupling}{2}(1+e^{\sfj z_\ell})\right].
	\end{equation}
	Then, the radiated power constraint is the summation of the radiated power from all DMA subarrays as
	\begin{equation}
		P_k(\bFrak, \bFanak, \bFdigk) = \sum_{\ps=0}^{\Pps-1} \Pink ( 1- |S_{12,\ps}^{\mathsf{all}} |^2).
	\end{equation}
	While this provides a simple and accurate method to calculate the radiated power for a tri-hybrid DMA array, a more detailed model that includes return loss can also be found in \cite{carlson2024analysisa}. 
	
	The DMA radiated power constraint couples all three precoding layers. The radiated power depends on the reconfigurable antenna precoder $\bFrak$ through the configurations $z_m$, and it depends on the analog $\bFanak$ and digital $\bFdigk$ precoders through the input power $\Pink$. This interdependence complicates the optimization because the reconfiguration mechanism directly affects the power budget. Capturing this coupling is important because it reveals how the forward scattering behavior of the waveguide links the reconfigurable antenna, analog, and digital precoders.
	
	
	DMAs trade per-element active driving for a waveguide-fed aperture whose slot responses are tuned by shifting each element’s resonance. This enables electronically steered, directive beams while naturally supporting tri-hybrid implementations where multiple subarrays are controlled through a smaller number of phase shifters and RF chains. The key drawback is that the Lorentzian constraint couples the achievable magnitude and phase, and the effective weights are additionally shaped by forward scattering and waveguide propagation. Thus, the element weights are not independently configurable in the same way as a conventional array. Overall, DMAs offer a tradeoff between hardware cost and beamforming capabilities, but realizing their full potential requires novel precoder algorithms that explicitly account for the constrained, propagation-coupled element responses. Electromagnetic reconfiguration transforms the wireless channel from a fixed entity into a controllable function of the precoder, altering the signal processing problem.

	\subsection{Polarization reconfigurable antennas}\label{subsec: polarization reconfigurable}

	Polarization reconfigurable arrays adapt the polarization of outgoing and incoming signals to mitigate polarization mismatches. The polarization of an antenna refers to the orientation of the electric field oscillations with respect to a given two-dimensional coordinate basis. Commonly used polarizations for communication systems include vertical polarization, horizontal polarization, slant polarization and right-handed circular polarization and left-handed circular polarization. Losses from channel depolarization and transmit/receive polarization mismatches are generally addressed with the use of dual-polarized antennas, which come at the cost of needing two separate antennas and feeds per one dual-polarized array element. In contrast, polarization reconfigurable antennas can be fed with a single RF chain to reduce the feed network complexity and the system power consumption.
	
	The polarization of an antenna depends on the relationship between the horizontal and vertical components of its complex gain pattern. For a given antenna, let $g_{\sfV}$ denote the vertical component of the complex gain pattern and $g_{\sfH}$ denote the horizontal component of the complex gain pattern, where the directional dependence of the patterns is omitted for convenience. The complex gain pattern vector is then $\bv = [g_{\sfH} \, g_{\sfV}]^{\rmT}$. To see this, we decompose the gain vector in terms of a polarization angle $\theta \in [0, 2\pi)$, horizontal phase offset $\psi_{\sfH} \in [0, 2\pi)$ and vertical phase offset $\psi_{\sfV} \in [0, 2\pi)$ as
	\begin{equation}
		\bv = \norm{\bv} \left[\cos(\theta) e^{\sfj \psi_{\sfH}}  \, \, \sin(\theta) e^{\sfj \psi_{\sfV}} \right]^{\rmT}.
	\end{equation}
	The polarization can then be characterized through $\theta$ and the phase offset difference $\psi = \psi_{\sfV} - \psi_{\sfH}$. Setting $\theta = \pi/2$ and $\psi = 0$ gives a vertically polarized field, while taking $\theta = \pi/4$ and $\psi = -\pi/2$ gives right-handed circular polarization. The polarization of the radiated field varies with direction around the antenna, but it is typically constant over the main lobe of the gain pattern.
	
	Reconfigurable antennas can dynamically adapt the polarization angle and phase offset to generate different polarization states. Let $\gamma(\theta, \psi) \in [0,1]$ denote the reconfiguration efficiency of the antenna and $g_0$ be a direction-dependent constant that accounts for the gain pattern norm and the phase of the horizontal component. The reconfigured gain pattern can be written in terms of the reconfigurable polarization vector $\bp(\theta, \psi) = \sqrt{\gamma}(\theta, \psi) \left[ \cos(\theta) \, \sin(\theta)e^{\sfj \psi} \right] ^{\rmT}$ as
	\begin{equation}
		\label{eq:gain_pol_vector}
		\bv(\theta, \psi) = g_0 \, \bp(\theta, \psi).
	\end{equation}
	The gain pattern constant can be absorbed into the channel, meaning $\bp(\theta, \psi)$ contains all of the relevant reconfiguration information. The norm of $\bp(\theta, \psi)$ is only dependent on the reconfiguration efficiency. Lossless reconfiguration would imply that $\gamma(\theta, \psi) = 1$. The polarization vector may depend on both direction and frequency. Let $\bphi$ denote the direction of propagation in terms of the azimuth and elevation angle. Modifying $\eqref{eq:gain_pol_vector}$ to account for $\bphi$ and the subcarrier index gives
	\begin{equation}
		\label{eq:gain_pattern_pol_vector}
		\bv_k(\bphi, \theta, \psi) = g_{0,k}(\bphi) \bp(\theta_k(\bphi), \psi_k(\bphi)).
	\end{equation}
	The selected tuning for the reconfigurable antenna determines $\theta_k(\bphi)$ and $\psi_k(\bphi)$ and $\gamma_k(\theta(\bphi), \psi(\bphi))$.
	
	We characterize the effect of polarization reconfiguration on the tri-hybrid MIMO through polarization precoders and combiners that act on an unpolarized channel. For the $n_{\sfT}$ antenna and $k$th subcarrier, let $\bp_{\sfT, k,n_{\sfT}}$ be the reconfigured polarization vector. Let $\bp_{\sfR, k,n_{\sfR}}$ be defined analogously for the receiver. We define the $2 \NT \times \NT$ reconfigurable precoder
	\begin{equation}
		\bFrak = \mathrm{blkdiag}\left(\bp_{\sfT, k,1}, \, \ldots, \, \bp_{\sfT, k,\NT} \right)
	\end{equation}
	and reconfigurable combiner
	\begin{equation}
		\bW_{\mathsf{ra},k} = \mathrm{blkdiag}\left(\bp_{\sfR, k,1}, \, \ldots, \, \bp_{\sfR, k,\NR} \right).
	\end{equation}
	These matrices completely characterize the polarization states of the transmit and receive arrays. The antennas in an array may all have the same polarization or they may all have distinct settings.
	Since the transmit and receive polarizations both affect the channel, the effective channel is a function of both the polarization precoder and combiner, i.e., 
	$\bH_{\mathsf{eff},k}(\mathbf{F}_{\mathsf{ra},k}, \bW_{\mathsf{ra},k})$. 
	
	The input-output model can be simplified under the assumption that the polarization state is not directionally dependent. For the $k$th subcarrier $n_{\sfT}$th transmit antenna and $n_{\sfR}$th receive antenna, we denote the channel from transmit polarization $X$ to receive polarization $Y$ as $H^{Y_{} , X_{}}_{k, n_{\sfR}, n_{\sfT}}$. We then define the \emph{unpolarized} channel for antenna pair ($n_{\sfR}$, $n_{\sfT}$) as
	\begin{equation}
		\bH_{\mathsf{up}, k, n_{\sfR}, n_{\sfT}} = \begin{bmatrix}
			H^{\sfH , \sfH}_{k, n_{\sfR}, n_{\sfT}} & H^{\sfH, \sfV}_{k, n_{\sfR}, n_{\sfT}} \\
			H^{\sfV, \sfH}_{k, n_{\sfR}, n_{\sfT}} & H^{\sfV, \sfV}_{k, n_{\sfR}, n_{\sfT}}
		\end{bmatrix}.
	\end{equation}
	If the polarization of each antenna is not directionally dependent, then the effective channel for antenna pair ($n_{\sfR}$, $n_{\sfT}$) is
	\begin{equation}
		H_{\mathsf{eff},k, n_{\sfR}, n_{\sfT}} =\bp_{\sfR, k, n_{\sfR}}^* \bH_{\mathsf{up}, k, n_{\sfR}, n_{\sfT}} \bp_{\sfT, k, n_{\sfT}}.
	\end{equation}
	The $2\NR \times 2\NT$ MIMO unpolarized channel $\bH_{\mathsf{up},k}$ is a block matrix with $2 \times 2$ submatrices $\bH_{\mathsf{up}, k, n_{\sfR}, n_{\sfT}}$. The effective channel is then
	\begin{equation}
		\label{eq_pol_recon_eff_channel}
		\bH_{\mathsf{eff}, k} =  \bW_{\mathsf{ra},k}^* \bH_{\mathsf{up},k} \bFrak.
	\end{equation}
	While the reconfigurable combiner can be absorbed into the effective channel for static receivers, the form in \eqref{eq_pol_recon_eff_channel} illustrates that both the transmit and receive polarization must be included.
	
	
	Polarization reconfiguration indirectly affects the radiated power through the reconfiguration efficiency. Let $\gamma_{n,k}(\bFrak)$ denote the reconfiguration efficiency of the $n$th transmit antennas at subcarrier $k$. Similarly to the case of pixel antennas, we define a diagonal efficiency matrix
	\begin{equation}
		\bm{\Gamma}(\bFrak) = \text{diag} (\gamma_{1,k}(\bFrak), \, \ldots, \, \gamma_{\NT,k}(\bFrak)).
	\end{equation}
	The transmit power is then
	\begin{equation}
		\label{eq_pol_recon_power}
		P_k\!(\bFrak, \bFanak, \bFdigk)\!\!=\! \trace{\mathbf{F}_{\mathsf{dig},k}^* \mathbf{F}_{\mathsf{ana},k}^* \bm{\Gamma}(\bFrak)  \bFanak \bFdigk }\!\!.
	\end{equation}
	Assuming all polarization configurations have the same efficiency $\gamma_k$, the transmit power can be simplified as
	\begin{equation}
		\label{eq_pol_recon_power_scalar_eff}
		P_k(\bFrak, \bFanak, \bFdigk) = \gamma_k \trace{ \mathbf{F}_{\mathsf{dig},k}^* \mathbf{F}_{\mathsf{ana},k}^*\bFanak \bFdigk }.
	\end{equation}
	In both \eqref{eq_pol_recon_power} and \eqref{eq_pol_recon_power_scalar_eff}, the effect of $\bFrak$ is determined by the reconfiguration efficiency because the polarization precoder is expressed in terms of two orthogonal polarizations.

	\begin{figure}
		\centering
		\includegraphics[width=0.45\textwidth]{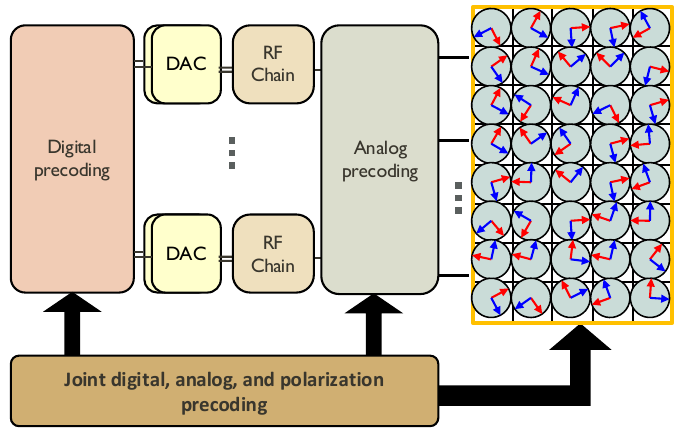}
		\caption{System model for the polarization-reconfigurable tri-hybrid architecture. Each antenna independently adjusts its polarization state, parameterized by an angle and phase offset between two orthogonal components. Unlike architectures that emphasize spatial pattern steering (e.g., parasitic, pixel/FAS, and DMA tri-hybrid models treated earlier in this section), polarization reconfiguration does not reshape the beam pattern but instead adapts the field orientation to match the channel polarization.}
		\label{fig: pol recon system model}
	\end{figure}
	
	

	

	
	\subsection{Stacked intelligent metasurfaces}\label{subsec: SIM}
	
	A stacked intelligent metasurface (SIM) is a transmitter architecture
	that performs beamforming in the wave domain by passing radiated signals
	through multiple programmable metasurface
	layers~\cite{intro_stacked_1}.
	The SIM consists of $L$ metasurface layers, each comprising
	$N_\mathsf{m}$ meta-atoms, placed in front of $M$ transmit antennas.
	Each meta-atom applies a tunable phase shift to the local
	electromagnetic field, typically controlled by a varactor or PIN diode.
	As the radiated wave propagates through the stack, it is successively
	transformed by each layer, and the cascade of per-layer
	transformations realizes spatial multiplexing and beamforming in the
	wave domain.
	Because a significant portion of the spatial processing is offloaded
	to the passive metasurface stack, the SIM can relax the requirements
	on the digital and analog precoding layers.
	In a tri-hybrid configuration, the digital and analog layers
	complement the wave-domain processing by providing per-subcarrier
	adaptivity and additional spatial selectivity that the
	frequency-flat SIM cannot achieve alone. 
	
	Physically, the wave is radiated from the antennas and impinges on the SIM layers. Each layer applies a tunable phase shift at each meta-atom position. The shifted wavefront then propagates to the next layer that applies a separate tunable phase shift. The phase shifting process is applied at every layer for $L$ layers and the cascaded effect is compounded onto the wavefront. The cascade effect achieves a spatial precoding operation through the successive wave interactions. The SIM adds minimal power overhead relative to digital or analog precoding stages since the SIM is passive. 
	
	We describe the input-output model for the SIM-based tri-hybrid
	architecture and connect it to the generic formulation
	in~\eqref{eqn: generic input output system model}.
	Consider $M$ transmit antennas placed behind an $L$-layer SIM with
	$N_\mathsf{m}$ meta-atoms per layer.
	The $\Nrf$ RF chains connect to the $M$ antenna feeds through an
	analog precoding network
	$\bFanak \in \bbC^{M \times \Nrf}$, and the digital precoder is
	$\bFdigk \in \bbC^{\Nrf \times \NS}$.

	We model the SIM using multiport network theory and transfer
	scattering parameters
	(T-parameters)~\cite{yahya2025tparam}.
	The cascaded architecture of SIM makes T-parameters a natural
	choice, since the overall transfer matrix of the stack is obtained by
	direct multiplication of the individual layer and propagation
	matrices. In contrast, the S-parameter recursion~\cite{nerini2024physical}
	requires nested matrix inversions that grow with the number of
	layers, while the Z-parameter model~\cite{abrardo2025multiport}
	used for single-layer parasitic modeling in the previous subsection
	leads to convoluted block-matrix expressions.
	Each metasurface layer~$\ell$ is then characterized as a reciprocal surface with
	per-element transmission and reflection
	coefficients~$T_n^{[\ell]}$ and~$R_n^{[\ell]}$, for
	$n = 1, \ldots, N_\mathsf{m}$.
	\begin{equation}\label{eqn: SIM G blocks}
		\begin{bmatrix}
			[\bG^{[\ell]}_{11}]_{n,n} & [\bG^{[\ell]}_{12}]_{n,n} \\[4pt]
			[\bG^{[\ell]}_{21}]_{n,n} & [\bG^{[\ell]}_{22}]_{n,n}
		\end{bmatrix}
		=
		\begin{bmatrix}
			T_n^{[\ell]} - \dfrac{(R_n^{[\ell]})^2}{T_n^{[\ell]}}
			& \dfrac{R_n^{[\ell]}}{T_n^{[\ell]}} \\[6pt]
			-\dfrac{R_n^{[\ell]}}{T_n^{[\ell]}}
			& \dfrac{1}{T_n^{[\ell]}}
		\end{bmatrix}.
	\end{equation}
	The off-diagonal blocks $\bG^{[\ell]}_{12}$ and $\bG^{[\ell]}_{21}$
	capture the effect of reflections at each layer and vanish when $R_n^{[\ell]} = 0$.
	
	\begin{figure}
		\centering
		\includegraphics[width=0.45\textwidth]{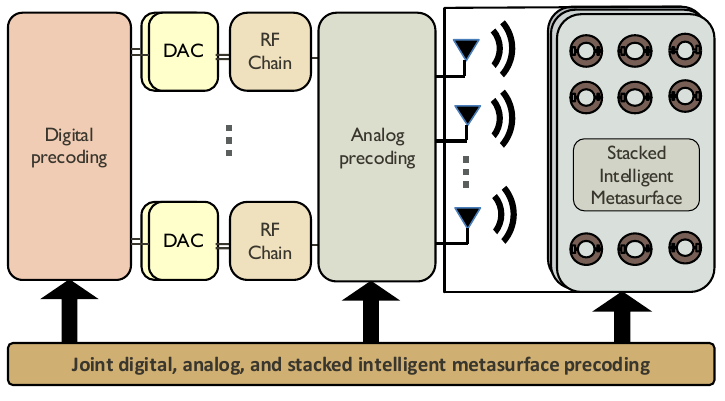}
		\caption{System model for the SIM-based tri-hybrid architecture. The $M$ transmit antennas radiate into a stack of $L$ metasurface layers, each containing $N_\mathsf{m}$ tunable meta-atoms that apply per-element phase shifts to the propagating wavefront. An analog precoding network connects the $\Nrf$ RF chains to the $M$ antenna feeds. The cascaded layers perform spatial processing in the wave domain, reducing the burden on the digital and analog precoding stages.}
		\label{fig: SIM system model}
	\end{figure}
	The electromagnetic propagation between consecutive layers is
	captured by the T-parameter matrix
	$\bP^{[\ell]} \in \bbC^{2N_\mathsf{m} \times 2N_\mathsf{m}}$, which
	can be computed from the impedance matrix of the interlayer transmission
	medium~\cite{yahya2025tparam}
	or approximated using the Rayleigh--Sommerfeld~(RS) diffraction
	kernel~\cite{nassirpour2025sumrate}. The interlayer propagation scattering parameters are captured by $\bG^{[\ell]}$ after every layer $\ell$ and the scattering parameters in between $\ell$ and $\ell+1$ is captured by $\bP^{[\ell]}$. 
	The end-to-end T-parameter matrix of the $L$-layer SIM is obtained by
	cascading all layer and propagation matrices
	as~\cite{yahya2025tparam}
	\begin{equation}\label{eqn: SIM cascade tparam}
		\bT_I = \bG^{[1]}\, \bP^{[1]}\, \bG^{[2]}\, \bP^{[2]}
		\cdots \bP^{[L-1]}\, \bG^{[L]},
	\end{equation}
	where $\bT_I \in \bbC^{2N_\mathsf{m} \times 2N_\mathsf{m}}$ with
	$(i,j)$th block $\bT_{I,ij} \in \bbC^{N_\mathsf{m} \times N_\mathsf{m}}$.
	A key advantage of the T-parameter formulation is that the cascade
	in~\eqref{eqn: SIM cascade tparam} involves only matrix
	multiplications, unlike the S-parameter formulation which requires
	nested matrix inversions that grow with~$L$~\cite{nerini2024physical}.

	The end-to-end transmission response of the SIM is
	$\bPsi = (\bT_{I,22})^{-1}$~\cite{yahya2025tparam}.
	We define the SIM reconfigurable antenna precoding matrix as
	$\bFrak = \bPsi \in \bbC^{N_\mathsf{m} \times N_\mathsf{m}}$. The SIM configuration~$\bFrak$ is frequency-flat, i.e., the
	same~$\bPsi$ applies across all $K$ subcarriers, since the
	meta-atom tuning states do not vary with frequency.
	Let $\bH_{\mathsf{IT}} \in \bbC^{N_\mathsf{m} \times M}$ denote the
	channel from the $M$ transmit antennas to the first SIM layer,
	which is determined by the fixed geometry. 
	Let $\bH_{\mathsf{RI},k} \in \bbC^{\NR \times N_\mathsf{m}}$ denote
	the wireless propagation channel from the last SIM layer to the
	$\NR$ receive antennas at the $k$th subcarrier.
	The effective MIMO channel in~\eqref{eqn: generic input output
		system model} is then
	\begin{equation}\label{eqn: effective MIMO channel SIM}
		\bH_{\mathsf{eff},k}(\bFrak) =
		\bH_{\mathsf{RI},k}\, \bFrak\, \bH_{\mathsf{IT}}.
	\end{equation}
	A distinguishing feature of this model is that the reconfigurable
	antenna precoder~$\bFrak$ is embedded between two channel
	matrices rather than multiplying the channel from one side.
	For a conventional antenna array without a SIM,
	$\bFrak = \bI_{N_\mathsf{m}}$ and the effective channel
	reduces to the fixed matrix
	$\bH_{\mathsf{RI},k}\, \bH_{\mathsf{IT}}$.

	When the meta-atoms are lossless and reflectionless,
	i.e., $R_n^{[\ell]} = 0$ and $|T_n^{[\ell]}| = 1$ for all~$n, \ell$,
	the off-diagonal blocks of~$\bG^{[\ell]}$ vanish and the layer
	T-parameter matrix simplifies
	to~\cite{yahya2025tparam}
	\begin{equation}\label{eqn: SIM ideal G}
		\bG^{[\ell]} = \mathrm{blkdiag}\!\left(
		\bTheta^{[\ell]},\;
		(\bTheta^{[\ell]})^{-1}
		\right),
	\end{equation}
	where $\bTheta^{[\ell]} = \diagg(
	e^{\sfj\theta_1^{[\ell]}},\, \ldots,\,
	e^{\sfj\theta_{N_\mathsf{m}}^{[\ell]}})$
	collects the unit-modulus phase shifts on layer~$\ell$.
	In this case, the end-to-end response~$\bPsi$ reduces to the cascade of
	phase-shift matrices and interlayer propagation matrices used in the
	simplified SIM channel models~\cite{nassirpour2025sumrate, intro_stacked_1}.
	The optimization variables are the $L \times N_\mathsf{m}$ phase
	shifts $\{\theta_n^{[\ell]}\}$, each constrained to
	$[0, 2\pi)$, and the feasible set for the reconfigurable antenna
	precoder in~\eqref{eqn: antenna precoder} is
	\begin{equation}\label{eqn: SIM feasible set}
		\cF_{\mathsf{ra}} = \left\{
		\bPsi = (\bT_{I,22})^{-1}
		~:~ \theta_n^{[\ell]} \in [0,2\pi),
		\; \forall\, \ell, n
		\right\},
	\end{equation}
	where $\bT_I$ depends on $\{\theta_n^{[\ell]}\}$ through the
	cascade in~\eqref{eqn: SIM cascade tparam}.

	The radiated power at the $k$th subcarrier is
	\begin{equation}\label{eqn: SIM radiated power}
		P_k(\bFrak, \bFanak, \bFdigk) =
		\| \bFrak\, \bH_{\mathsf{IT}}\, \bFanak\, \bFdigk \|^2_\Fnorm,
	\end{equation}
	which couples the SIM configuration to the power constraint
	through the end-to-end transmission response~$\bFrak = \bPsi$.
	This expression is general and compatible with any meta-atom
	model that produces per-element transmission and reflection
	coefficients~\cite{nerini2024physical, yahya2025tparam}.
	Under the ideal assumptions of lossless meta-atoms
	($R_n^{[\ell]} = 0$, $|T_n^{[\ell]}| = 1$) and
	power-conserving interlayer propagation, the SIM precoder
	becomes unitary and the radiated power reduces to
	$\| \bFanak \bFdigk \|^2_\Fnorm$, decoupling the SIM
	phase shifts from the power constraint. This reduction requires that $\bH_{\mathsf{IT}}^\ast \bH_{\mathsf{IT}} = \bI_M$, as such,
	the transmit antennas couple without losses into the first
	metasurface layer. When this condition is not met, the
	general expression in~\eqref{eqn: SIM radiated power}
	applies.

	A SIM trades digital and analog hardware complexity for passive
	wave-domain processing, which introduces several design tradeoffs.
	Increasing~$L$ provides additional beamforming degrees of freedom, but large cascades of~\eqref{eqn: SIM cascade tparam} become increasingly non-convex. Each added layer introduces $N_\mathsf{m}$ phase variables whose effect on the objective is modulated by all subsequent layers.
	The SIM provides only phase control at each meta-atom, unless active amplifiers are integrated; any amplitude shaping must be
	provided by the digital or analog layers.
	Additionally, the frequency-flat nature of~$\bPsi$ limits the SIM
	in wideband OFDM systems, motivating the use of the per-subcarrier
	digital precoder~$\bFdigk$ and the analog precoding network to
	compensate.
	This presents a design dimension that is unique to
	SIM-based architectures: the allocation of beamforming responsibility
	between the digital-to-analog converter (DAC) resolution,
	the analog precoding network, and the SIM depth~$L$.
	With fewer DAC bits or a simpler analog network, the digital and
	analog layers have limited spatial selectivity, requiring a deeper
	SIM to compensate. Conversely, a higher-resolution digital backend
	relaxes the requirement on the number of SIM layers.
	Overall, realizing the full potential of the SIM in a tri-hybrid
	architecture requires joint optimization algorithms that account for
	the cascade structure, the frequency-flat constraint, and the
	interplay across all three precoding layers.
	
	\subsection{Pinching antennas}\label{subsec: pinching antennas}
\begin{figure}[!t]
		\centering
		\includegraphics[width=0.45\textwidth]{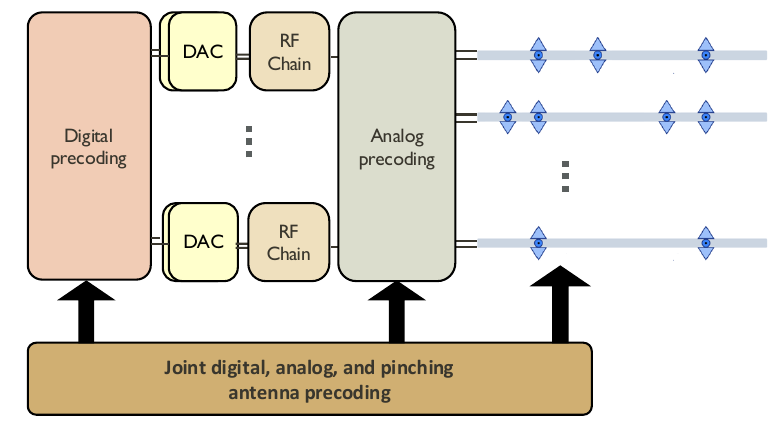}
		\caption{System model for the PASS-based tri-hybrid precoder. Each of the $P$ waveguides is equipped with $M_p$ movable pinches that couple guided-wave energy into free space at controllable positions. The sequential coupling creates a physical dependency between pinch locations and radiated amplitude, distinguishing this architecture from conventional arrays with independent element weights.}
		\label{fig:PASS-system-model}
	\end{figure}

	Pinching-antenna systems (PASS) introduce a new class of reconfigurable transceivers that interface electromagnetic waves between guided media and free space through spatially distributed coupling units named as ``pinching'' elements \cite{liu2026pinching}. In a PASS-aided tri-hybrid MIMO system with $P$ waveguides depicted in Fig. \ref{fig:PASS-system-model}, each waveguide $p \in \{1, \ldots, P\}$ is equipped with $M_p$ movable pinches, resulting in a total of $N_{\mathrm{pin}} = \sum\limits_{p=1}^P M_p$ pinching antenna elements as depicted in Fig. \ref{fig:PASS-system-model}.
	
	The reconfigurability depends not only on the absolute pinching positions but also on their relative positioning. This is because the pinching elements exhibit a \textit{unidirectional sequential coupling} where the power available for radiation at any given pinch is the residual energy of the guided wave after being attenuated by all preceding pinching elements. This interaction is characterized by the local amplitude coupling coefficient for the $m$-th pinch, $\delta_m \in (0, 1)$, defined as the ratio of the radiated field amplitude to the incident field amplitude.
	
	The reconfigurable precoding matrix $\mathbf{F}_{\text{ra},k} = \text{diag}(\mathbf{f}_{1,k}, \dots, \mathbf{f}_{P,k}) \in \mathbb{C}^{N_{\mathrm{pin}} \times P}$ is defined by the vectors $\mathbf{f}_{p,k} \in \mathbb{C}^{M_p \times 1}$, which represent the weights of the pinching elements. This pinching model is governed by the coupling coefficient $\delta_m \in (0,1)$ of the $m$th pinch which represents how hard the pinch is pressing on the waveguide. More precisely, when a dielectric pinch is applied to the waveguide, the interaction between the guided mode and the antenna mode yields a continuous exchange of energy over the physical length of the pinch. The strength of this interaction is dictated by the overlap of the transverse field distributions between the waveguide and the pinching element. The coupling coefficient $\delta_m$ defines the exact fraction of the electromagnetic field that has transitioned from the waveguide into the antenna structure at the moment it reaches the open end to radiate into free space. For this reason, the coupling coefficient $\delta_m$ determines the amplitude of the radiated field $\alpha_m$, which can be written due to the nature of the sequential pinching coupling along waveguides as \cite{wang2025modeling}
	\begin{equation}\label{eq:alpham}
		\alpha_m=\delta_m \left( \prod_{i=1}^{m-1} \sqrt{1 - \delta_i^2} \right).
	\end{equation}
	Here, the term $\sqrt{1 - \delta_i^2}$ represents the fraction of signal amplitude inside the waveguide after passing the $i$th pinch.
	
	
	By letting $\beta_g$ and $x_{p,m}$ denote the waveguide propagation constant and the physical position $x_{p,m}$ of the $m$-th pinch on waveguide $p$, the reconfigurable precoding is not only a function of the magnitude of the radiated field $\alpha_m$ but also depends on the complex exponential $e^{-\sfj \beta_g x_{p,m}}$ to account for the wave propagation distance along the waveguide until it reaches the $m$th pinching position. The $m$-th entry of the reconfigurable precoding matrix $\mathbf{F}_{\text{ra},k}$ at subcarrier $k$ is then expressed as \cite{wang2025modeling}
	\begin{equation}
		[\mathbf{f}_{p,k}]_m = \alpha_m \,  e^{-\sfj \beta_g x_{p,m}}.
	\end{equation}
	It is worth noting how the multiplicative nature of the amplitude decay creates a non-linear physical dependency between the beamforming weights of the elements sharing the same waveguide. This distinguishes the PASS model from conventional arrays. The effective wireless channel for the PASS-based tri-hybrid architecture is then the combination of the wireless channel $\bH_k$ and the PASS beamforming matrix $\bFrak$, i.e.,  $\bH_{\mathsf{eff},k}(\bFrak) = \bH_k \bFrak$. In contrast to DMAs where elements are fixed and propagation delays are treated as static channel properties, the mechanical mobility of pinching elements in PASS makes the propagation phase a directly reconfigurable degree of freedom \cite{ding2025analytical}. By keeping the phase term $e^{-\sfj\beta_g x_{p,m}}$ inside the beamforming vector $\mathbf{f}_{p,k}$, the model explicitly captures how the physical optimization of positions $\{x_{p,m}\}$ actively shapes the antenna response rather than merely reflecting a fixed hardware geometry.
	
	To characterize the coupling behavior in multi-antenna waveguides, we distinguish between two power modeling approaches \cite{wang2025modeling}. In the \textit{equal power model}, the length of the $m$th pinch is individually adjusted such that every antenna radiates an identical proportion of the total power. In other words, the radiated field has a uniform amplitude $\alpha$ for all $m$. While this model serves as a useful theoretical performance upper-bound and ensures balanced efficiency, it requires manufacturing each antenna with a unique physical length, which increases hardware complexity and cost. Conversely, the \textit{proportional power model} assumes that each pinching antenna is manufactured with a uniform length, resulting in an identical coupling coefficient $\delta$ for all elements. Although this leads to a sequential amplitude decay as given in (\ref{eq:alpham}), this approach is highly practical for large-scale deployment as it significantly reduces costs through uniform hardware production, as all pinching elements have the same length.
	\begin{figure}[t]
		\centering
		\hspace*{-0.2cm}
		\includegraphics[scale=0.32]{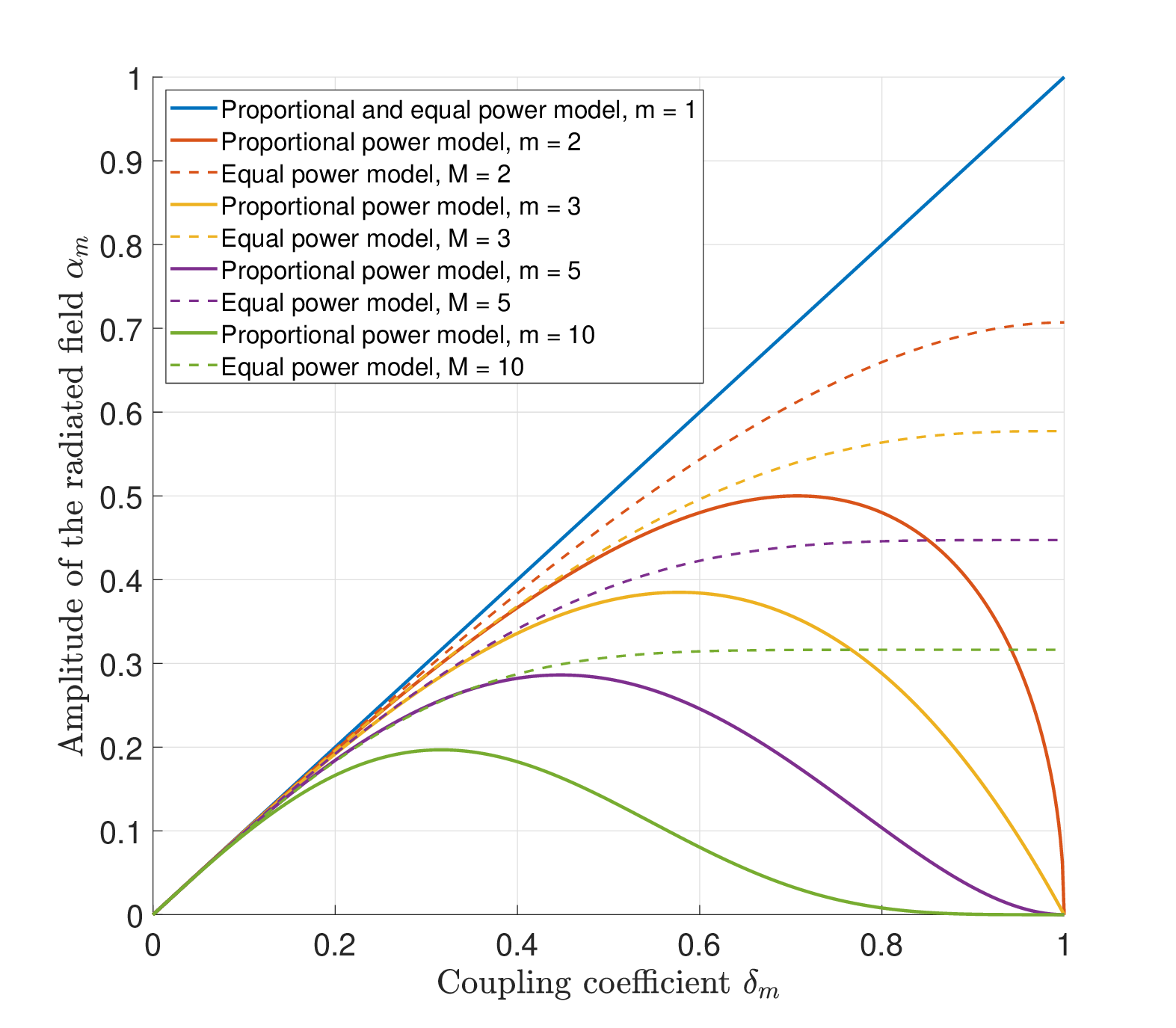} 
		\caption{Impact of coupling coefficient $\delta_m$ on the amplitude of the radiated field $\alpha_m$. Solid lines represent the proportional power model for various pinching indices $m$, while dashed lines indicate the equivalent amplitude required for an equal power model.}
		\label{fig:delta-alpha-impact}
	\end{figure}

	Fig. \ref{fig:delta-alpha-impact} depicts the amplitude of the radiated field $\alpha_m$ against the coupling coefficient $\delta_m$ where the signal energy is sequentially distributed across multiple pinching elements along a waveguide. The proportional power model (in solid lines) characterizes a system where all pinching antennas share a uniform physical length and hence the same coupling coefficient $\delta$. For the first element (i.e., $m=1$), the amplitude increases linearly with $\delta$.  For subsequent elements (i.e., $m>1$), however,  the amplitude degrades beyond a specific optimal coupling coefficient. While increasing $\delta$ initially increases the amplitude, it eventually exhausts the guided wave's power, causing the amplitude for later elements to decay toward zero. This model is highly practical for reducing hardware manufacturing costs due to the use of uniform hardware. In contrast, the equal power model (in dashed lines) represents a theoretical scenario where each pinch's length is individually adjusted to ensure every antenna radiates an identical proportion of the total power. This results in a uniform amplitude $\alpha_m$ that plateaus as $\delta$ increases, providing a performance benchmark for balanced waveguide gains. Although this model serves as an upper-bound for analysis, it requires unique physical lengths for each antenna, which increases hardware complexity and cost.
	
	The waveguide extracts energy sequentially along its length, so a portion of the input signal reaches the end of the guide without radiating. To prevent reflections, a matched terminal load typically absorbs this residual energy. Consequently, the total power radiated by waveguide $p$, denoted as $\eta_p$, is expressed as
	\begin{equation}
		\eta_p = \sum_{m=1}^{M_p} |\alpha_m|^2 = 1 - \prod_{m=1}^{M_p} (1 - \delta_m^2).
	\end{equation}
	We therefore introduce $\mathbf{\Gamma} = \text{diag}(\eta_1, \dots, \eta_P)$ to effectively filter out the residual power that is not radiated but is instead absorbed by the matched loads. This allows us to define the total radiated power $P_k$ as the energy successfully coupled into free space instead of the total transmit power radiated supplied to the waveguides. The total radiated power $P_k$ can be expressed as
	\begin{equation}\label{eq:PASS-Pk}
		P_k(\bFrak, \bFanak, \bFdigk) = \mathsf{Tr}(\bF^\ast_{\mathsf{dig},k} \bF^\ast_{\mathsf{ana},k}\,\mathbf{\Gamma}\,\bFanak\bFdigk).
	\end{equation}
	The power model in (\ref{eq:PASS-Pk}) is invariant to the physical positions $\{x_{p,m}\}$ of the pinching elements as radiation efficiency depends solely on the coupling coefficients $\{\delta_m\}$. This creates a decoupled design methodology where beam direction can be optimized through antenna positioning without impacting the total power budget.
	
	\subsection{Non-radiating wires}\label{subsec: non-radiating wires}
	
	Non-radiating wires are specialized leaky-wave antennas suited for near-field communication by providing a controlled pathway for electromagnetic waves \cite{martin1984leaky}. Unlike traditional antennas that rely on free-space radiation, these structures can operate in both the reactive and radiative near-field regions. They provide several compelling advantages including enhanced security due to the short communication range and precise device positioning, low susceptibility to external interference, and low environmental footprint achieved by mitigating wasteful radio emission and RF pollution \cite{GLine,martin1970radiocommunication}. While PASS offer high reconfigurability by adjusting the pinching positions, non-radiating wires are deployed with static positions of physical feeds. Yet, they retain its versatility through load-based reconfigurability that regulates the power exchange between these discrete feeding ports.
	
\begin{figure}[t]
		\centering
		\includegraphics[width=0.45\textwidth]{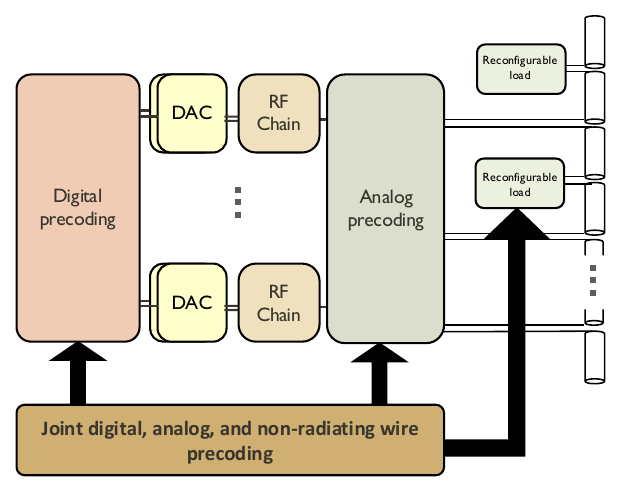}
		\caption{Connected dipole model for the non-radiating wire. Periodic feeding points spaced $\Delta$ apart along a thin wire of radius $a$ drive the global current distribution. Each feed acts as an infinitesimal gap whose impedance depends on the local excitation voltage and the mutual coupling with all other ports. Reconfigurable load impedances at non-excited ports shape the electromagnetic field pattern and the inter-port power exchange.}
		\label{fig:non-rad-wire}
	\end{figure}

	A non-radiating wire can be modeled as a one-dimensional connected array of dipoles with periodic feeding points distributed along a thin wire with radius $a$ and uniform periodic spacing $\Delta$ as depicted in Fig. \ref{fig:non-rad-wire}. The wave propagation within this wire is characterized by the free-space wave number $k_0$ and the intrinsic impedance $Z_0$.
	
	Each feed point acts as an infinitesimal gap that drives the global current distribution. To describe the field in the wave-number domain, we first define $\alpha$ as the continuous longitudinal wave number, which serves as the spatial frequency variable. In the presence of discrete feeds, we define $\alpha_{\ell} = \alpha - \frac{2\pi \ell}{\Delta}$ as the wave number associated with the $\ell$-th spatial harmonic. It is then possible to define the transverse propagation constant of the $\ell$-th mode $\beta_{\ell} = \sqrt{k_0^2 - \alpha_{\ell}^2}$, which determines the radial decay or oscillation of that specific mode. 	To characterize the physical feeding mechanism, it has been shown in \cite{akrout2026guidedwirelesstechnologynearfield} that the impedance at the $m$-th excitation point $z[m\Delta]$ can be expressed by separating the spatial-harmonic Green's-function sum from the feed-phasing integral. Let $\Phi(\alpha)$ collect the per-mode radial response (with $\beta_{\ell}$ depending on $\alpha$ through $\alpha_{\ell}$ as above). Then
	\begin{equation}\label{eq:wire-impedance}
		\begin{aligned}
			\Phi(\alpha) &= \sum_{\ell=-\infty}^{\infty}\frac{1}{\beta_{\ell}^{2}J_{0}(\beta_{\ell}a)H_{0}^{(2)}(\beta_{\ell}a)},\\
			z[m\Delta] &= \frac{Z_{0}\Delta^{2}}{8\pi k_0}\int_{0}^{\frac{2\pi}{\Delta}}\frac{e^{j\alpha m\Delta}}{\Phi(\alpha)}\,d\alpha,
		\end{aligned}
	\end{equation}
	where $J_0(\cdot)$ is the zero-order Bessel function and $H_0^{(2)}(\cdot)$ is the zero-order Hankel function of the second kind. This model reveals that the current at any feed point $n\Delta$ is not determined solely by the local voltage $V_n$. Instead, it is a result of both local excitation and all induced currents propagated along the wire, as embodied by the kernel $\Phi(\alpha)$ in (\ref{eq:wire-impedance}). Since the impedance at discrete ports depends exclusively on the relative spacing $m\Delta$ between excitation points, the transmit impedance matrix $\mathbf{Z}_{\mathsf{ra}}$ is constructed such that its $(i, j)$-th entry is defined by the impedance in (\ref{eq:wire-impedance}) evaluated at the relative distance between ports, i.e., $[\mathbf{Z}_{\mathsf{ra}}]_{i,j} = z[|i - j|\cdot\Delta]$. Depending on which ports are being excited or loaded, entries corresponding to non-excited or open-circuited positions result in zero elements. Similarly to parsitic arrays, for a given loading configuration denoted as $\textrm{diag}(\mathbf{z}_{\mathsf{load}})$, the effective channel is given by $\bH_{\mathsf{eff},k}(\bFrak) = \bH_k \bFrak$ where $\bFrak = (\mathbf{Z}_{\mathsf{ra}} + \textrm{diag}(\mathbf{z}_{\mathsf{load}}) )^{-1}$.

	A key feature of the non-radiating wire is its inherent reconfigurability through impedance matching.  While a subset of feeding ports are actively excited, the remaining non-excited positions along the connected array can be terminated with specific load impedances. By adaptively terminating the non-excited positions with specific load impedances, the non-radiating wire can reconfigure its electromagnetic field pattern and the mutual coupling between feeding points. Concretely, the configurability of the loads alters the impedance matching at each port, which in turn changes the reflection coefficients and hence the amount of power exchanged between the feeding points along the wire. This global field/power interaction within the wire allows the non-radiating wire to act as a programmable medium without requiring physical alterations to its geometry, unlike in PASS.
	
\begin{figure}[t]
	\centering
	\begin{subfigure}[b]{0.5\textwidth}
		\hspace{-0.4cm}\includegraphics[scale=0.225]{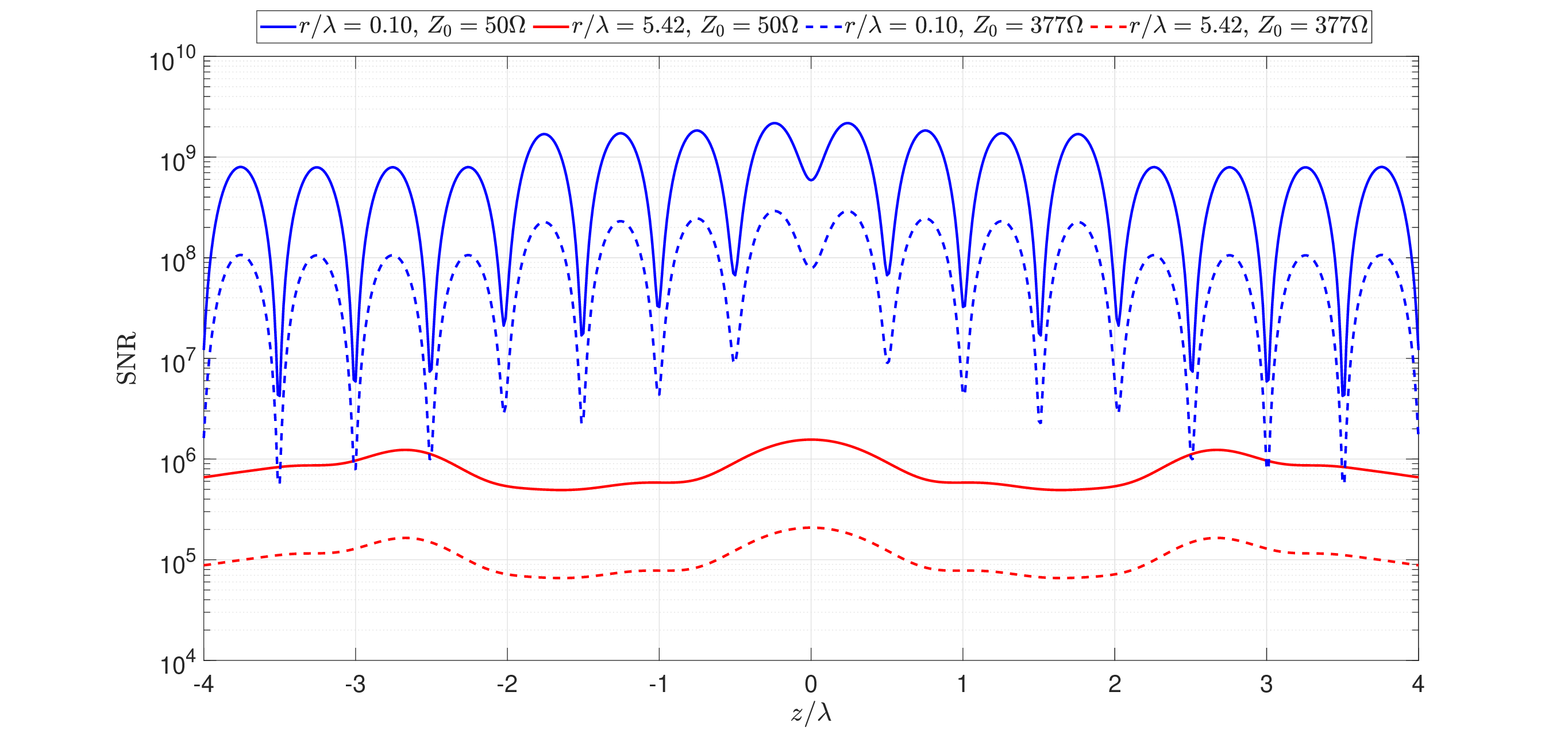}
		\caption{Single excitation port}
		\label{fig:1-ext-load-377}
	\end{subfigure}
	\hfill
	\begin{subfigure}[b]{0.5\textwidth}
		\hspace{-0.4cm}\includegraphics[scale=0.225]{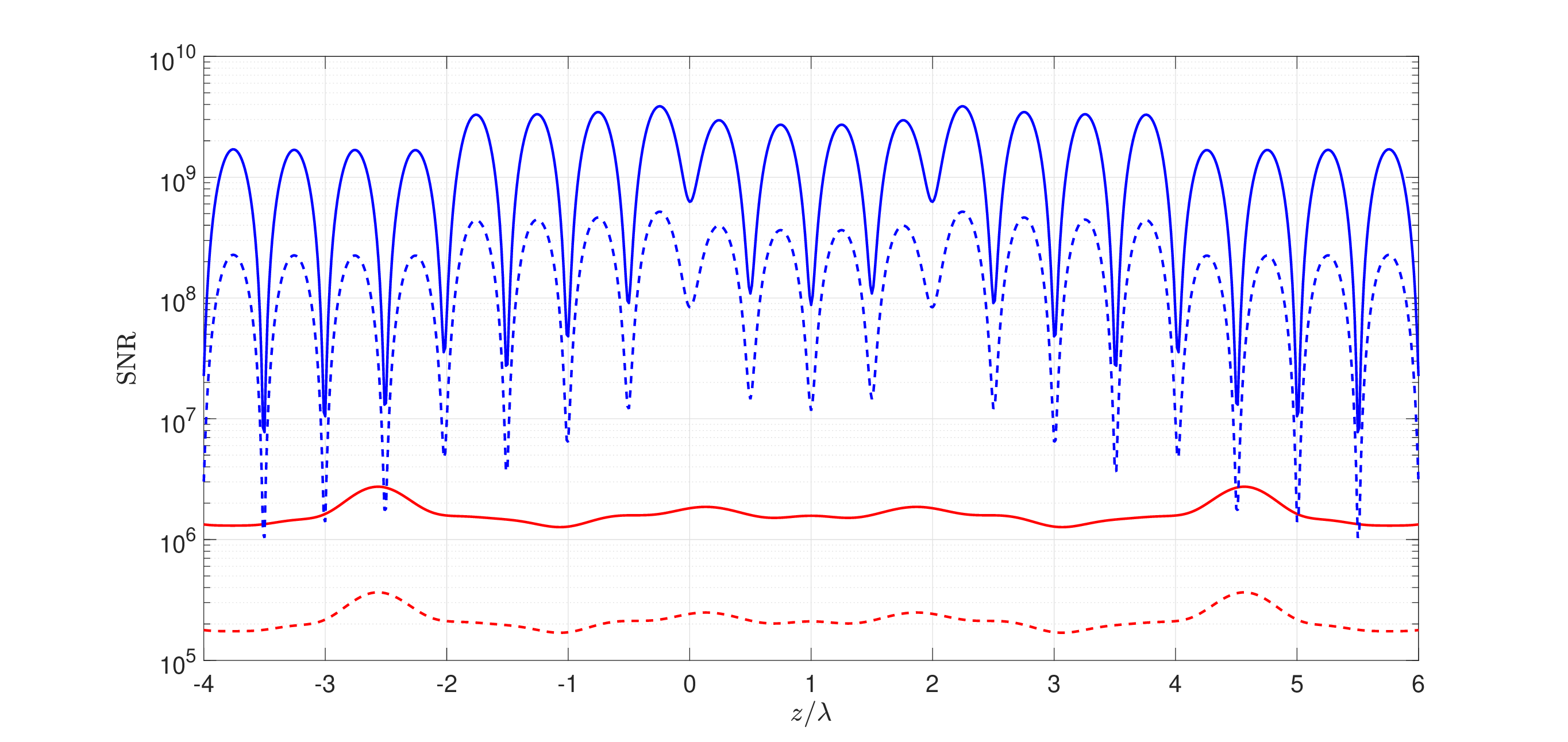}
		\caption{Two excitation ports}
		\label{fig:2-ext-load-377}
	\end{subfigure}
	\caption{SNR as a function of longitudinal position $z/\lambda$ and radial distance $r/\lambda$ for a non-radiating wire along the wire axis ($r=0$). Increasing the number of active feeds produces a more spatially continuous field coverage by reducing standing-wave nulls. Transitioning from a free-space $377\,\Omega$ load to a $50\,\Omega$ termination shifts the standing-wave pattern, confirming that impedance mismatch at non-excited ports steers the reactive energy profile.}
	\label{fig:non-radiating-channel}
\end{figure}

	Fig. \ref{fig:non-radiating-channel} shows signal-to-noise ratio (SNR) as a function of longitudinal position $z/\lambda$ and radial distance $r/\lambda$ for a non-radiating wire along the axis $r=0$. Comparing Fig.~\ref{fig:non-radiating-channel} (a) with one excitation port to Fig.~\ref{fig:non-radiating-channel} (b)  with two excitation ports shows that increasing the number of active ports enables a more spatially continuous field coverage by minimizing the channel fluctuations. Moreover, the transition from a free space $377\,\Omega$ load  to a $50\,\Omega$ load  significantly alters the reflection coefficients at the non-excited ports, which yields change in the position of standing wave peaks and nulls to ensure uniform spectral efficiency for a mobile user. This confirms how impedance mismatching can be an effective tool to steer the reactive energy profile. Overall, this demonstrates that the non-radiating wire can be treated as a fixed-pinch version of PASS, where electronic load modulation replaces mechanical movement to achieve spatial field tuning.
	

	\begin{table*}[t]
		\centering
		\caption{Comparison of reconfigurable antenna architectures}
		\label{tab:antenna_properties}
		\small
		\renewcommand{\arraystretch}{1.35}
		\setlength{\tabcolsep}{4.5pt}
		
		\begin{tabularx}{\textwidth}{@{}l *{5}{>{\setlength{\hsize}{0.93\hsize}\raggedright\arraybackslash}X} >{\setlength{\hsize}{1.35\hsize}\raggedright\arraybackslash}X@{}}
			\toprule
			\textbf{Type} & 
			\textbf{Reconfiguration principle} & 
			\textbf{Tuning variable} & 
			\textbf{EM control} & 
			\textbf{Key constraint} & 
			\textbf{Tradeoff} \\
			\midrule
			
			Parasitic arrays 
			& Mutual coupling via tunable parasitic loads 
			& Load reactances 
			& Induced currents; beamforming weights 
			& Coupled magnitude–phase response 
			& Reduced hardware, limited beamforming freedom \\
			
			Pixel arrays / FAS 
			& Aperture/position selection 
			& State or port index 
			& Active aperture or radiator location 
			& Discrete state space; efficiency variation 
			& Low RF complexity, combinatorial design \\
			
			Dynamic metasurface antennas 
			& Resonance tuning of waveguide-fed slots 
			& Slot state $z_m$ 
			& Slot response; effective aperture weights 
			& Lorentzian forward-scattering coupling 
			& Low power, constrained beam control \\
			
			Polarization-reconfigurable antennas 
			& Polarization-state adaptation 
			& Polarization angle/phase 
			& Polarization vector; matching 
			& Efficiency loss; Tx/Rx coupling 
			& Reduced mismatch, limited spatial control \\
			
			Stacked intelligent metasurfaces 
			& Cascaded wave-domain processing 
			& Per-layer meta-atom phase 
			& End-to-end transmission 
			& Multilayer non-convex coupling 
			& Passive gain, harder optimization \\
			
			Pinching antennas 
			& Guided-to-radiated coupling (movable pinches) 
			& Pinch position/coupling 
			& Extraction point, amplitude, phase 
			& Sequential waveguide coupling 
			& Strong spatial control, limited independence \\
			
			Non-radiating wires 
			& Load-controlled field shaping 
			& Terminal/load impedance 
			& Standing-wave profile; reactive field 
			& Global impedance coupling 
			& Near-field shaping, termination-sensitive \\
			
			\bottomrule
		\end{tabularx}
	\end{table*}

	The radiated power of a non-radiating wire is related to the port voltage vector $\mathbf{v}_{\mathsf{ra}}$ and the real part of its admittance $\cR\{\mathbf{Z}^{-1}_{\mathsf{ra}}\}$. At a carrier $k$, the voltages fed to the wire $\mathbf{v}_{\mathsf{ra}}$ are generated by analog and digital precoders $\mathbf{F}_{\mathsf{ana},k}$ and $\mathbf{F}_{\mathsf{dig},k}$ acting on a vector of transmit voltage $\mathbf{v}_{\mathsf{T}}$ such that $\mathbf{v}_{\mathsf{ra}} = \mathbf{F}_{\mathsf{ana},k}\,\mathbf{F}_{\mathsf{dig},k} \mathbf{v}_{\mathsf{T}}$. Assuming that each feeding point is assigned an equal normalized average power, i.e., $\mathbb{E}[\mathbf{v}_{\mathsf{T}} \mathbf{v}_{\mathsf{T}}^*] = \mathbf{I}$, the average radiated power at the $k$th subcarrier is given by
	\begin{equation}\label{eq:power-non-radiating-wire}
		\begin{aligned}
			&P_k(\bFrak, \bFanak, \bFdigk)\\
			&= \mathbb{E} \big[ \mathbf{v}_{\mathsf{ra}}^* \,\cR\{\mathbf{Z}^{-1}_{\mathsf{ra},k}\} \,\mathbf{v}_{\mathsf{ra}} \big]\\
			& = \mathsf{Tr}(\bF^\ast_{\mathsf{dig},k} \bF^\ast_{\mathsf{ana},k}\cR\{\mathbf{Z}^{-1}_{\mathsf{ra},k}\}\bFanak\bFdigk)\text{.}
		\end{aligned}
	\end{equation}
	By optimizing the digital, analog, and antenna precoders subject to the average power in (\ref{eq:power-non-radiating-wire}), the system can adaptively allocate power to mitigate standing wave nulls in the reactive field. This control over standing waves provides a reconfigurable advantage for secure near-field communication.
	
	\subsection{Summary}\label{subsec: summary reconfig antennas}
	
	
	The reconfigurable antenna architectures in this section can be distinguished by how they introduce electromagnetic programmability into the tri-hybrid MIMO architecture. Parasitic arrays and non-radiating wires rely on load-based interactions, pixel arrays and FAS rely on discrete aperture or position selection, polarization-reconfigurable antennas adapt the field polarization state, while DMAs, SIMs, and pinching antennas exploit guided or wave-domain propagation to shape the radiated response.
	
	Despite these different hardware realizations, several common themes emerge. Pixel arrays/ FAS, and parasitic arrays both reduce the burden on active RF hardware, but the former operate through discrete state selection whereas the latter exploit mutual coupling through tunable loads. Likewise, the coupled magnitude-phase behavior in parasitic arrays is closely related to the Lorentzian constraint in DMAs, since in both cases the achievable antenna response is restricted by electromagnetic feasibility rather than being freely assigned as in a fully active phased array.
	
	Waveguide-mediated architectures reveal another important connection. DMAs and pinching antennas both operate on a shared guiding structure, so the response of one radiating element depends on the propagation and extraction behavior of the others. DMAs, however, use fixed slot locations with electronic resonance tuning, whereas pinching antennas reconfigure the spatial extraction points along the waveguide. Non-radiating wires are closely related in that they also shape fields through interactions along a connected guiding structure, but they do so through electronic load control rather than mechanical repositioning. Table~\ref{tab:antenna_properties} summarizes these similarities and differences.
	These cross-architecture tradeoffs motivate the need for a common system-level metric that can compare the benefits and costs of reconfiguration across different designs.


	
	
	
	
	\section{Efficiency of reconfigurability: A new figure of merit for the tri-hybrid MIMO architecture}\label{sec: Efficiency of reconfigurability: A new figure of merit for the tri-hybrid MIMO architecture}
	
	This section introduces the reconfigurability efficiency factor (REF), a system-level metric for evaluating reconfigurable antenna designs. We first motivate the need for such a metric by reviewing the limitations of conventional antenna figures of merit. We then define the REF and discuss its interpretation and robustness.

	\begin{table*}[ht]
		\centering
		\caption{Summary of conventional antenna metrics and their limitations for reconfigurable systems}
		\label{tab: conventional metrics for antennas}
		\renewcommand{\arraystretch}{1.45}
		\setlength{\tabcolsep}{5pt}
		\begin{tabularx}{\textwidth}{@{}l c >{\raggedright\arraybackslash}X >{\raggedright\arraybackslash}X >{\raggedright\arraybackslash}X@{}}
			\toprule
			\textbf{Metric} & \textbf{Symbol} & \textbf{Math definition} & \textbf{Physical significance} & \textbf{Limitation for reconfigurability} \\
			\midrule
			
			Directivity & $D$ & $\displaystyle \frac{U(\theta, \phi)}{P_{rad} / 4\pi}$ & Measures focusing capability relative to an isotropic source. & Ignores all losses (ohmic/dielectric) and mismatch; theoretical maximum only. \\
			
			Gain & $G$ & $\displaystyle \eta_{rad} \cdot D$ & Accounts for focusing and internal ohmic/dielectric losses. & Ignores impedance mismatch losses, which vary significantly across tuning states. \\
			
			Realized gain & $G_{real}$ & $\displaystyle (1 - |\Gamma|^2) G$ & The actual power gain delivered to the far-field from the source. & Does not account for the implementation cost (power/complexity) required to achieve the gain. \\
			
			Radiation efficiency & $\eta_{rad}$ & $\displaystyle P_{rad} / P_{in}$ & Ratio of power radiated to power accepted by the antenna. & Often degrades in reconfigurable designs due to switch resistance; does not capture system-level benefit. \\
			
			Aperture efficiency & $\epsilon_{ap}$ & $\displaystyle \frac{G \lambda^2}{4\pi A_{phys}}$ & Measures how effectively the physical area is utilized. & Static concept; fails to quantify "virtual" aperture expansion via parasitic coupling. \\
			
			Envelope correlation & $\rho_e$ & $\displaystyle \approx \frac{\bigl|\iint \vec{F}_1 \cdot \vec{F}_2^{\ast}\bigr|^{2}}{\bigl(\iint |\vec{F}_1|^{2}\bigr)\bigl(\iint |\vec{F}_2|^{2}\bigr)}$ & Measures statistical independence of radiation patterns (diversity). & Ensures states are distinct, but does not guarantee that any state provides a net performance gain (SNR). \\
			
			Active reflection & TARC & $\displaystyle \sqrt{\frac{\sum |b_i|^2}{\sum |a_i|^2}}$ & Characterizes matching efficiency in active arrays with coupling. & Complex to measure across all combinations of reconfigurable states; ignores hardware cost. \\
			
			\bottomrule
		\end{tabularx}
	\end{table*}

	\subsection{Need for a new metric}\label{subsec: need new metric}
	Standard antenna metrics such as directivity ($D$), realized gain ($G_\mathsf{realized}$), and radiation efficiency ($\eta_\mathsf{rad}$) were developed to characterize static apertures. We review the definitions of conventional metrics used in antennas and arrays literature in Table~\ref{tab: conventional metrics for antennas}.
	While these metrics effectively quantify the electromagnetic performance of a fixed structure, they fail to capture the specific value proposition of reconfigurable antennas, which is the ability to adapt dynamically based on the propagation channel. For reconfigurable systems, metrics like the envelope correlation coefficient (ECC) are often used to ensure pattern diversity. This metric, however, only measures the statistical independence of the radiation patterns; it does not account for the cost of achieving this diversity.
	
\begin{figure}[t]
		\centering
		\includegraphics[width=\columnwidth]{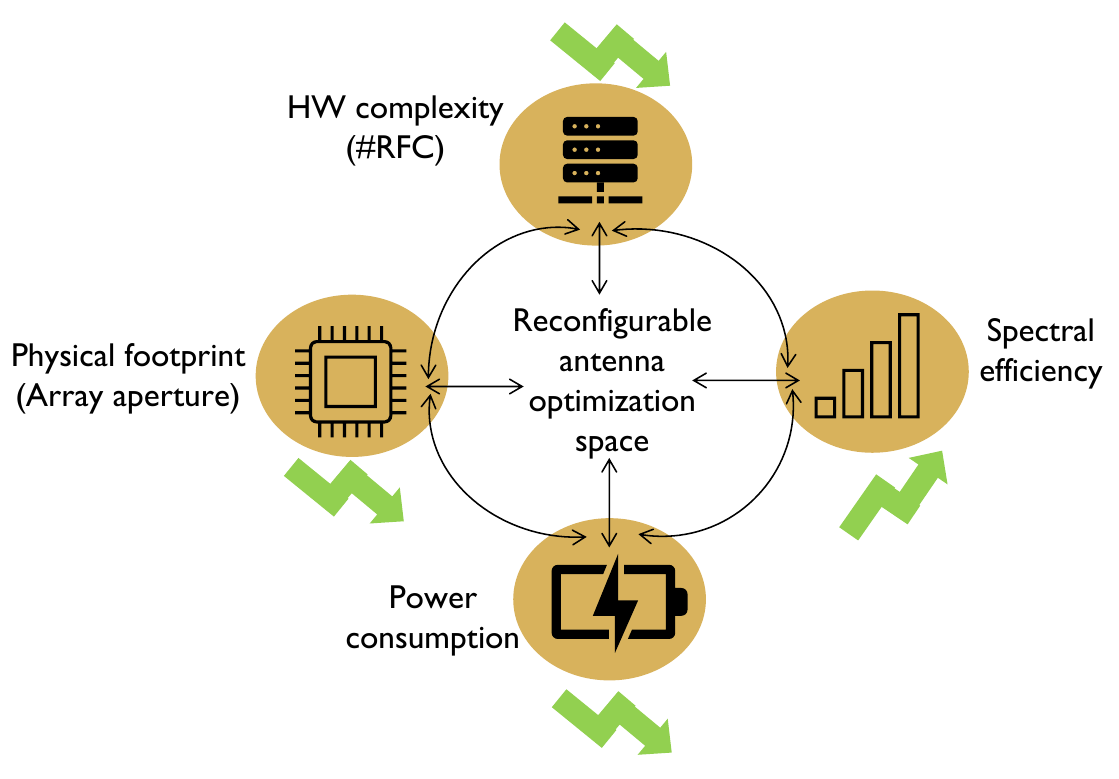} 
		\caption{Design space for reconfigurable antenna optimization, defined by four competing objectives: spectral efficiency, hardware complexity, physical footprint, and power consumption. Conventional metrics capture only one dimension of this space. The reconfigurability efficiency factor (REF) proposed in Section~\ref{subsec: metric definition} jointly accounts for gains and costs across these dimensions.}
		\label{fig:tradeoff_diagram}
	\end{figure}

	To fully evaluate a reconfigurable wireless system, one must consider multiple interdependent design objectives simultaneously. As illustrated in Fig.~\ref{fig:tradeoff_diagram}, the design space is defined by four  nodes: spectral efficiency, hardware complexity, power consumption, and physical footprint. Conventional metrics typically focus on only one of these nodes in isolation. A reconfigurable architecture, however, often improves one node while incurring a penalty in another.

	
	Consequently, there is a lack of a unified figure of merit that quantifies the system-level benefit of reconfigurability normalized by its implementation cost. Without such a metric, comparing a reconfigurable tri-hybrid MIMO architecture to a conventional fully-digital array is challenging without properly accounting for tradeoffs in different metrics. To rigorously justify the adoption of  tri-hybrid MIMO architectures, we require a metric that relates the relative change in system performance  to the relative change in system complexity or power.

	\subsection{Metric definition}\label{subsec: metric definition}
	We propose a generalized system-level metric, termed the \textit{reconfigurability efficiency factor} (REF), denoted as $\Upsilon$. This metric quantifies the net benefit of reconfigurability by aggregating the weighted improvements in system objectives against the weighted penalties incurred.
	
	Let $\mathcal{M} = \{Q_1, Q_2, \dots, Q_N\}$ be the set of $N$ relevant system metrics such as spectral efficiency, total power consumption, hardware complexity, and physical aperture. We first define the normalized relative change $\Delta(Q_i)$ for any metric $Q_i \in \mathcal{M}$ between the reconfigurable system and a static reference as
	\begin{equation}
		\Delta(Q_i) = \left| \frac{Q_{i, \mathsf{reconfig}} - Q_{i, \mathsf{static}}}{Q_{i, \mathsf{static}}} \right|.
	\end{equation}
	To evaluate a specific design, we partition the varying metrics into two disjoint subsets based on the design intent. The set $\mathcal{G}$ contains metrics exhibiting a \textit{desirable} trend, representing the benefits. The set $\mathcal{B}$ contains metrics exhibiting an \textit{undesirable} trend, representing the costs or penalties. Any metric $Q_k$ held constant during the comparison belongs to the constraint set $\mathcal{K}$ where the relative change is approximately zero.
	
	To account for the varying relevance of different system objectives, we assign a weight $w_k \in [0, 1]$ to each metric $Q_k$, subject to the constraint that the sum of all weights used in the evaluation equals one. These weights allow the metric to be tailored to specific applications; for instance, a system designer might assign a higher weight to power consumption for Internet of Things (IoT) devices compared to base station applications.
	
	The generalized REF is then defined as the ratio of the weighted sum of benefits to the weighted sum of costs
	\begin{equation}\label{eqn: REF}
		\Upsilon = \frac{\sum_{i \in \mathcal{G}} w_i \cdot \Delta(Q_i)}{\sum_{j \in \mathcal{B}} w_j \cdot \Delta(Q_j)} \Bigg|_{\mathcal{K}}.
	\end{equation}
	Because the metric depends on the specific choice of benefit set $\mathcal{G}$, cost set $\mathcal{B}$, and constraint set $\mathcal{K}$, REF values are directly comparable only when these assignments are defined consistently. Consequently, REF values computed using different benefit/cost metrics or different reference conditions should be interpreted within their own design study rather than compared numerically across different reconfigurable antenna types. In particular, the numerical instantiations in Section~\ref{sec: numerical examples ref} assign different physical quantities to the benefit and cost sides of the ratio. For example, transmit power may appear as a benefit in one study (power saved at fixed spectral efficiency) but as a cost in another (hardware power to realize a gain). So a larger REF for one antenna type does not indicate a uniformly preferable architecture relative to a smaller REF reported elsewhere unless the same $\mathcal{G}$, $\mathcal{B}$, $\mathcal{K}$, baseline, and operating point are enforced by construction.
	
	\paragraph{Performance-limited regime}
	When the primary goal is maximizing a performance metric $Q_\mathsf{perf}$ such as spectral efficiency despite an increase in a cost metric $Q_\mathsf{cost}$ like hardware complexity, we set the benefit set $\mathcal{G}$ to contain only the performance metric and the cost set $\mathcal{B}$ to contain only the cost metric. The metric simplifies to
	\begin{equation}
		\Upsilon_\mathsf{perf} = \frac{w_1 \Delta(Q_\mathsf{perf})}{w_2 \Delta(Q_\mathsf{cost})} \propto \frac{\Delta(Q_\mathsf{perf})}{\Delta(Q_\mathsf{cost})}.
	\end{equation}
	For example, if a reconfigurable tri-hybrid array achieves a 50\% increase in spectral efficiency ($\Delta(Q_\mathsf{perf}) = 0.5$) while incurring a 25\% increase in hardware complexity due to additional switches ($\Delta(Q_\mathsf{cost}) = 0.25$), the resulting factor is $\Upsilon_\mathsf{perf} = 2.0$. This value greater than unity indicates that the performance gain efficiently outweighs the added cost.
	
	\paragraph{Resource-limited regime}
	When the design intent is to reduce a resource metric $Q_\mathsf{save}$ such as power consumption while tolerating a degradation in a performance metric $Q_\mathsf{perf}$ like spectral efficiency, we set the benefit set $\mathcal{G}$ to contain the resource metric and the cost set $\mathcal{B}$ to contain the performance metric. The metric becomes
	\begin{equation}
		\Upsilon_\mathsf{save} = \frac{w_1 \Delta(Q_\mathsf{save})}{w_2 \Delta(Q_\mathsf{perf})} \propto \frac{\Delta(Q_\mathsf{save})}{\Delta(Q_\mathsf{perf})}.
	\end{equation}
	Consider a battery-operated IoT sensor that utilizes a simplified parasitic antenna to reduce power consumption by 80\% ($\Delta(Q_\mathsf{save}) = 0.8$) compared to a full MIMO array, while suffering only a 10\% drop in throughput ($\Delta(Q_\mathsf{perf}) = 0.1$). This results in a high efficiency factor of $\Upsilon_\mathsf{save} = 8.0$, demonstrating a highly favorable trade-off for energy-constrained applications.
	
	\paragraph{Form-factor limited regime}
	When the system has flexibility in physical size but is constrained by power or hardware budgets, the goal is to reduce power or cost by increasing the physical aperture while maintaining constant performance. In this regime, the benefit set $\mathcal{G}$ contains the resource saving $Q_\mathsf{save}$ and the cost set $\mathcal{B}$ contains the physical area $A_\mathsf{phys}$. The metric is defined as
	\begin{equation}
		\Upsilon_\mathsf{area} = \frac{w_1 \Delta(Q_\mathsf{save})}{w_2 \Delta(A_\mathsf{phys})} \Bigg|_{Q_\mathsf{perf}}.
	\end{equation}
	For instance, in a massive MIMO base station where space is abundant, expanding the aperture area by 20\% ($\Delta(A_\mathsf{phys}) = 0.2$) to reduce total power consumption by 40\% ($\Delta(Q_\mathsf{save}) = 0.4$) yields a factor of $\Upsilon_\mathsf{area} = 2.0$. This quantifies the effectiveness of increasing array aperture for energy savings.

	\subsection{Robustness and sensitivity considerations}\label{subsec: robustness sensitivity}
	To ensure the proposed REF metric is mathematically consistent and robust against scaling ambiguities, we impose two critical constraints on its calculation.
	
	\subsubsection{Linearity constraint}
	A common ambiguity in antenna metrics is the choice between logarithmic (e.g., dB, dBm) and linear scales. Calculating a relative percentage change on a logarithmic quantity leads to inconsistent results that depend on the reference value. Therefore, we define the inputs $Q$ strictly in their linear physical units.
	\begin{itemize}
		\item \textbf{Power ($P$):} Must be measured in Watts (W), not dBm.
		\item \textbf{Gain/Directivity ($G$):} Must be in linear magnitude (dimensionless ratio), not dBi.
		\item \textbf{Spectral Efficiency (SE):} Measured in bps/Hz (naturally linear).
		\item \textbf{Hardware Complexity:} Measured in absolute counts (e.g., number of elements/switches).
	\end{itemize}
	Using linear units ensures that the ratio $\Delta(Q)$ represents a true physical scaling of the system resources.
	
	\subsubsection{Singularity avoidance}
	The efficiency factor $\Upsilon$ is a ratio metric, which is susceptible to numerical instability if the denominator (the cost penalty) approaches zero. This scenario occurs when a reconfigurable design achieves performance gains with negligible added cost. While this represents an ideal engineering outcome (infinite efficiency), it can create an ``illusion'' of benefit if the denominator is dominated by measurement noise or quantization error.
	
	To mitigate this, the metric is valid only when the denominator represents a statistically significant change. We introduce a regularization threshold $\epsilon$, representing the minimum detectable change or engineering tolerance:
	\begin{equation}
		\Upsilon_\mathsf{robust} = \frac{\Delta(Q_\mathsf{benefit})}{\max(\Delta(Q_\mathsf{cost}), \epsilon)}.
	\end{equation}
	In practical evaluations, if $\Delta(Q_\mathsf{cost}) < \epsilon$, the trade-off is considered cost-neutral, and the designs should be compared based on absolute performance difference ($Q_\mathsf{benefit}$) rather than efficiency.

	\begin{figure}[t]
		\centering
		\includegraphics[width=1.05\columnwidth]{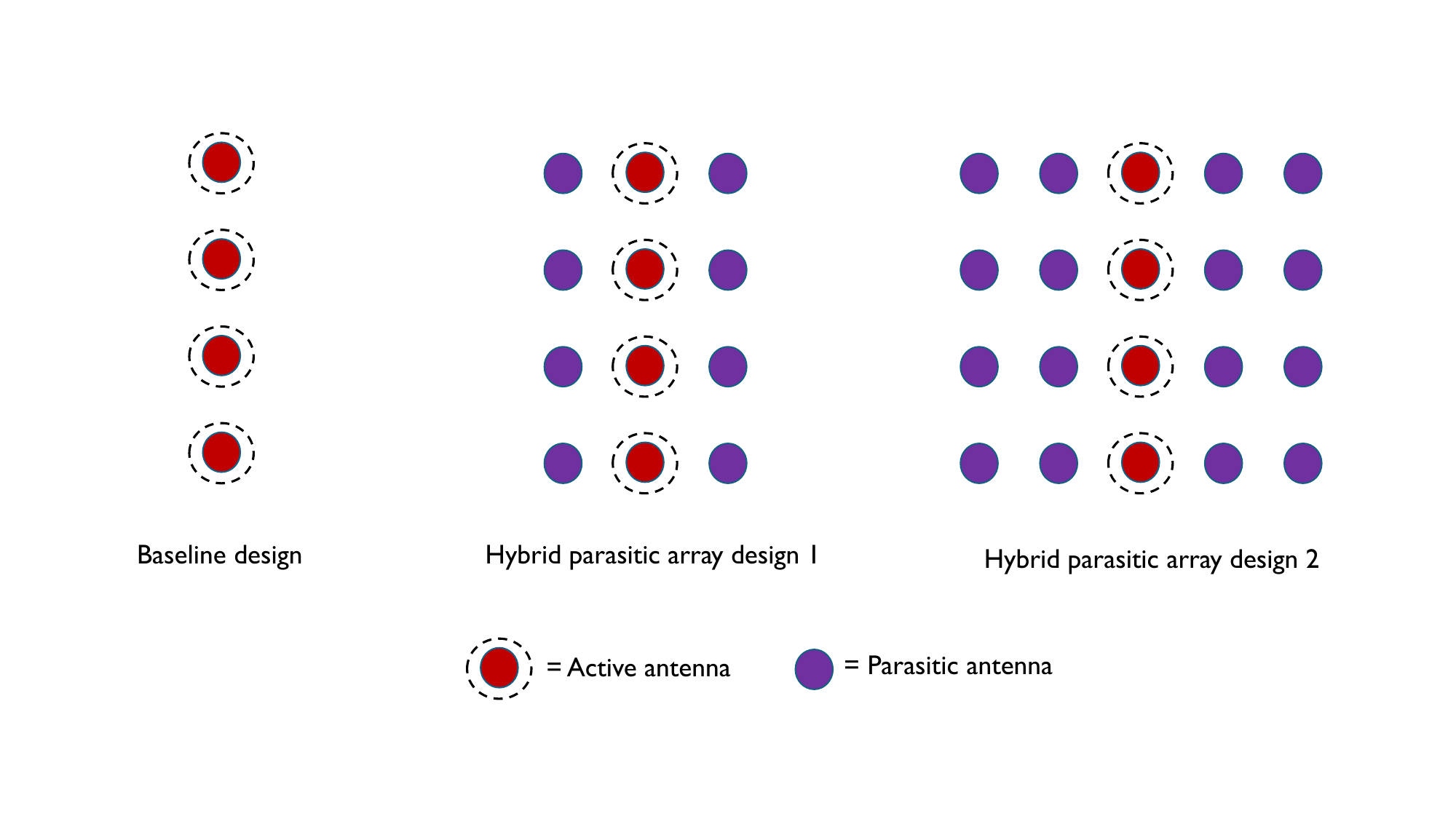} 
		\caption{Three parasitic array designs compared at fixed spectral efficiency: a baseline with $N_\sfA=4$ active antennas, a hybrid design with $N_\sfA=4$ active and $N_\sfP=8$ parasitic antennas (aperture $3A_1$), and a larger hybrid design with $N_\sfA=4$ active and $N_\sfP=16$ parasitic antennas (aperture $5A_1$). Adding parasitic elements reduces the required transmit power at the cost of increased aperture, and the REF quantifies which design trades aperture for power most efficiently.}
		\label{fig:parasitic_designs}
	\end{figure}
	
	\begin{table*}[ht]
		\centering
		\caption{Comparison of baseline and parasitic hybrid array designs using the proposed reconfigurability efficiency factor (REF)}
		\label{tab:design_comparison}
		\renewcommand{\arraystretch}{1.3} 
		\setlength{\tabcolsep}{8pt} 
		\begin{tabular}{l c c c c c c}
			\toprule
			\textbf{Design topology} & 
			\textbf{$A_\mathsf{phys}$} & 
			\textbf{$\Delta(A_\mathsf{phys})$} & 
			\textbf{$\mathsf{SE}$} & 
			\textbf{$P_\mathsf{tx}$} & 
			\textbf{$\Delta(P_\mathsf{tx})$} & 
			\textbf{REF ($\Upsilon$)} \\
			& (cm$^2$) & & (bps/Hz) & (mW) &  & (Dimensionless) \\
			\midrule
			
			\textbf{Baseline design} & 
			$A_1$ & 
			--- & 
			4.2 & 
			20 & 
			--- & 
			--- \\
			
			\textbf{Hybrid parasitic array design 1} & 
			$3 A_1$ & 
			2 & 
			4.22 & 
			12 & 
			-0.4 & 
			\textbf{0.2} \\
			
			\textbf{Hybrid parasitic array design 2} & 
			$5 A_1$ & 
			4 & 
			4.24 & 
			10.7 & 
			-0.465 & 
			\textbf{0.12} \\
			
			\bottomrule
		\end{tabular}
		\par\medskip
		\footnotesize{\textit{Note: The baseline design serves as the static reference ($Q_\mathsf{static}$) for all $\Delta$ calculations. $\Delta$ values represent normalized relative change. The REF is denoted $\Upsilon$, matching the notation in Section~\ref{subsec: metric definition}.}}
		\end{table*}

			\section{Numerical examples for the REF}\label{sec: numerical examples ref}
			
			To illustrate the usefulness of the REF metric, we provide numerical examples for different designs.  
			These examples are intended to show how REF can quantify tradeoffs within each antenna family, not to support direct numerical comparison across all families. In particular, the parasitic-array example uses transmit-power reduction versus aperture increase, the DMA example uses realized-gain improvement versus power consumption, and the SIM and polarization-reconfigurable examples use spectral-efficiency improvement versus power consumption. The same physical quantity can therefore appear in the numerator in one instantiation and in the denominator in another, and the units or scaling of $\Upsilon$ need not match (e.g., a dimensionless ratio versus a value reported in decibels). REF magnitudes from different subsections are therefore not interchangeable indices of overall antenna-class preference. Cross-architecture comparison would require deliberately fixing a shared $\mathcal{G}$, $\mathcal{B}$, $\mathcal{K}$, baseline, and operating point as in \eqref{eqn: REF}, which we do not pursue here.
			
			\subsection{Parasitic antenna arrays}\label{subsec: numerical parasitic}
			Let us assume the hybrid parasitic uniform planar array design from \cite{11241086}.  The baseline for comparison is a uniform linear array with active antennas.
			We compare the following three array designs as shown in Fig.~\ref{fig:parasitic_designs}.
			\begin{itemize}
				\item Baseline design: We assume a uniform linear array with all active antennas. Let $N_\sfA=4$. Let the aperture area be $A_{1}$.
				\item Hybrid parasitic array design 1:  We assume a uniform planar array with $N_\sfA=4$ active antennas and $N_\sfP=8$ parasitic antennas.  Let the aperture area be $A_{2}= 3 A_1$.
				\item Hybrid parasitic array design 2:  We assume a uniform planar array with $N_\sfA=4$ active antennas and $N_\sfP=16$ parasitic antennas.  Let the aperture area be $A_3=  5 A_1$.
			\end{itemize}
			We simulate the spectral efficiency for all three designs using the optimization algorithm and the simulation setup same as our prior work~\cite{11241086}. To interpret the tradeoff between transmit power and array aperture, we  compute spectral efficiency for different transmit powers and obtain the power required for each design to obtain the same spectral efficiency.  It is known that with an increase in the aperture, the beamforming gain increases which enables reduction in the transmit power to attain the same spectral efficiency at the receiver. Our goal is to understand which of the two parasitic designs is more efficient using the REF definition from \eqref{eqn: REF}.
			
			In Table~\ref{tab:design_comparison}, we tabulate the different metrics for the three designs and compare the REF of the two hybrid parasitic array designs. For the three designs, the spectral efficiency values are approximately the same.  As the transmit power reduces compared to the baseline design at the cost of increase in the array aperture, the REF is defined as $\Upsilon =\frac{|\Delta(P_\mathsf{tx})|}{|\Delta(A_\mathsf{phys})|}$.  We observe that design 1 outperforms design 2 in terms of the REF.  The two additional columns of parasitic elements in design 2 are spaced further apart from the active array central column. This reduces coupling effect and only provides an incremental gain.   This example shows that the choice of the design is critical in reconfigurable antennas. The REF metric is useful to compare two reconfigurable antenna designs and make a better choice in how to scale the array in an efficient way.

			\subsection{Dynamic metasurface antennas}\label{subsec: numerical dma}
			
			For DMAs, there is a crucial tradeoff when determining the size and number of elements in a tri-hybrid system. The varactor diodes and bias control consume very little power, thus it is often favorable to fit many elements on the same waveguide to improve the beamforming gain and spectral efficiency. Due to the power decay in the waveguide electromagnetic fields, however, only so many elements can be placed on the same waveguide before the beamforming gain benefits begin to diminish. Placing multiple waveguides together counteracts this effect, but each additional waveguide adds DACs and RF chains, which consume significantly more power than varactor diodes. We examine this tradeoff in terms of the REF metric to determine the optimal number of elements and waveguides for a DMA array.
			
			We assume the weight model described in Section \ref{sec: DMA} for the DMA elements. Waveguides are spaced along the $y$-direction, and DMA elements are spaced along the $x$-direction. We vary the number of DMA elements $\Nx$ and waveguides $\Ny$ to determine the best array design. Because the coupling strength $\coupling$ dictates the rate of power decay for the waveguide fields, we optimize $\coupling$ for each design based on the number of elements per waveguide $\Nsub$ such that the power of the fields at the end of the waveguide is $|S_{12}|^2\approx 0.1$ \cite{carlson2025wideband}. We also provide a limit to the coupling strength as $A_\mathsf{min} = 0.2$. This is due to physical constraints in the DMA element size, geometry, and waveguide height that may limit how weakly coupled each DMA element is with the waveguide. For each case, we assume a $\lambda/2$ spacing between elements and waveguide.
			
			To determine the fundamental size tradeoffs, our goal is to compare the performance gains of increasing the number of array elements with the larger hardware burden of power consumption. To do this, we use the model in \cite{RibeiroEtAlEnergyEfficiencyMmWaveMassive2018} to calculate the power consumed $\Pcons$ by the RF chains, DACs, and power amplifiers for each case. We also assume a DAC6578 as the DC biasing DAC for each varactor, which consumes around $0.65$mW of power per element. Next, we incorporate a genetic algorithm to configure the weights for the DMA elements to steer a beam pattern in a desired direction. We then define the REF as the ratio between the realized gain difference and the power consumption difference with $\Upsilon = \frac{|\Delta \Gre|}{|\Delta \Pcons|}$ to determine the size tradeoff.
			
\begin{figure}[t]
				\centering
				\IfFileExists{figures/dma_2d_gamma_ref.pdf}{\includegraphics[width=1\columnwidth]{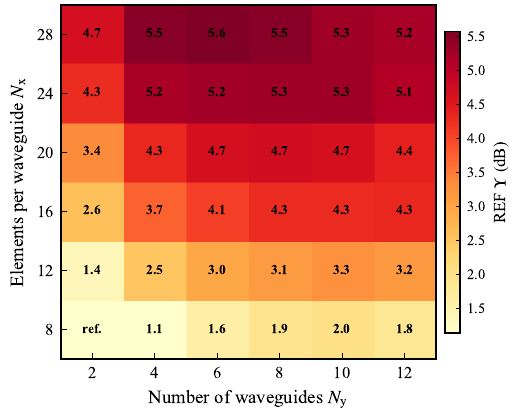}}{\includegraphics[width=1\columnwidth]{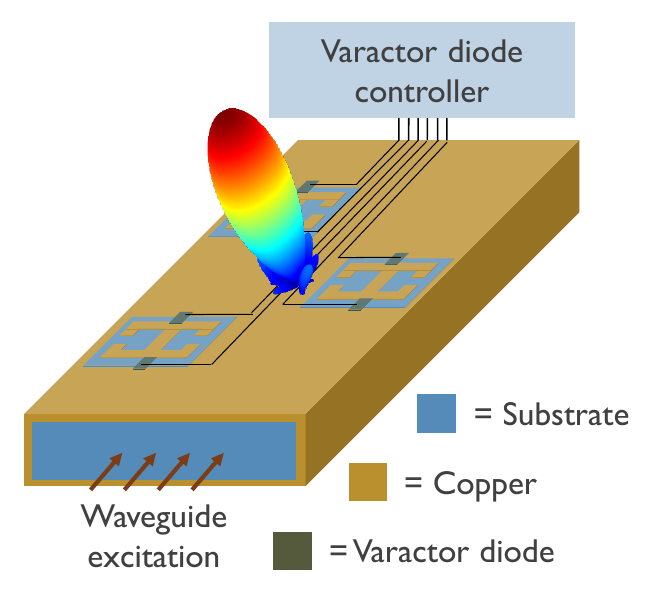}}
				\caption{REF for DMA arrays as a function of the number of elements per waveguide $\Nx$ and the number of waveguides $\Ny$, with baseline $\Nx=8$, $\Ny=2$ and $\lambda/2$ element and waveguide spacing. The coupling strength $\coupling$ is optimized per configuration so that approximately $10\%$ of the power remains at the waveguide end ($|S_{12}|^2\approx 0.1$). The REF peaks near $\Ny=6$ waveguides; beyond this point, the additional power consumed by DACs, RF chains, and varactor biasing outweighs the beamforming gain from extra waveguides.}
				\label{fig:dma_ref}
			\end{figure}

			Fig. \ref{fig:dma_ref} shows the simulation results for the REF for beam patterns steered to broadside. Each REF value is relative to the baseline case of $\Nx = 8, \Ny = 2$. First, we find that for a fixed $\Ny$, the REF improves as $\Nx$ increases. This is expected, since the low varactor biasing power consumption means that adding more elements per waveguide can increase the beamforming gain without a significant power consumption burden. It is also important to note that the REF improvements are more substantial when there are fewer $\Nx$ elements, since the power decay of the waveguide fields will begin to yield diminishing gain returns for a large number of elements. Moreover, we find that as $\Ny$ increases, the REF reaches a peak at $\Ny=6$ waveguides and then decreases for larger $\Ny$. This is due to the system power consumption as we add more waveguides, where the beamforming gain improvements do not outweigh the power consumption growth. Overall, the REF is shown to be critical here for finding the optimal DMA array size to balance high performance while maintaining a low hardware burden in terms of power consumption.
			
			\subsection{Stacked intelligent metasurfaces}\label{subsec: numerical sim}
			
			For the SIM, there is a crucial tradeoff between the number of metasurface layers
			and the DAC resolution at the digital backend.
			The SIM cascades $L$ layers of meta-atoms to shape the transmitted wavefront
			in the wave domain, as described in \secref{subsec: SIM}.
			It is typically favorable to add more layers since varactor power consumption  is minimal.
			The DAC power, however, scales exponentially with the number of bits. Therefore,
			a deeper SIM can compensate for coarse digital conversion at much lower power cost. Similar insights are available for dense arrays \cite{MezghaniEtAlMassiveMIMODenseArrays2020} and radar systems \cite{MazherEtAlImprovedCRBmmWaveRadar2021}.
			This section examines the tradeoff in terms of the REF to determine
			how many SIM layers are needed for a given DAC resolution.
			
			The simulation in this section considers a multiuser MIMO downlink with $M = 4$ transmit antennas,
			serving $K = 4$ single-antenna users, $L$ cascaded SIM layers
			with $N_\mathsf{m} = 64$ meta-atoms per layer, and $\Nrf = 4$ RF chains.
			The SIM phases are optimized to maximize the sum rate
			through the effective channel $\bH_{\mathsf{eff}} = \bH_{\mathsf{RI}} \bPsi \bH_{\mathsf{IT}}$. The transmitter applies quantized zero-forcing precoding with $1$ to $8$~DAC bits.
			The model in~\cite{RibeiroEtAlEnergyEfficiencyMmWaveMassive2018} is used to calculate
			the power consumed $\Pcons$ by the RF chains, DACs, and power amplifiers for each case.
			The DAC used for each varactor is a DAC6578,
			which consumes around $0.65$~mW of power per element. This is inline with the DMA analysis.
			The transmit power is $1$~W at $28$~GHz,
			and the results are averaged over $100$ channel realizations.
			
\begin{figure}[t]
				\centering
				\includegraphics[width=\columnwidth]{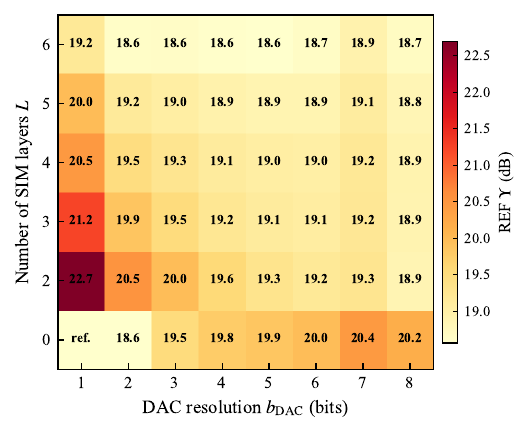}
				\caption{REF for SIM-aided quantized zero-forcing precoding
					as a function of SIM layers $L$ and DAC bits.
					Due to the tradeoff in DAC power consumption and wave-domain processing gain,
					there is an optimal combination of SIM depth and DAC resolution
					that provides the best spectral efficiency per unit power.}
				\label{fig:sim_ref}
			\end{figure}

			\figref{fig:sim_ref} shows the REF for different SIM configurations by layers $L$ and DAC bits.
			Each value is relative to the baseline case of $L = 0$ with $1$-bit DACs. The REF improved with additional layers for a given DAC resolution.
			This is expected, since the low varactor biasing power means
			that adding layers improves the effective channel quality
			without significant power burden.
			Notably, the REF improvements diminish
			for deeper SIM cascades
			since each additional layer yields smaller marginal gains in spectral efficiency.
			A SIM with $L = 6$ layers and $1$-bit DACs matches the performance
			of a system without SIM layers using $4$-bit DACs,
			while adding only $249.6$~mW ($6.3\%$ of the baseline power).
			The peak REF of $22.7$~dB occurs at $L = 2$ with $1$-bit DACs,
			where two SIM layers improve spectral efficiency by $4.9\times$
			with only a $2.1\%$ power increase.
			Overall, the REF demonstrates that the first few SIM layers
			provide the largest marginal gain.
			Moreover, the passive wave-domain processing can substitute for DAC resolution
			at roughly twice the power efficiency.
			
			\subsection{Polarization reconfigurable arrays}\label{subsec: numerical polarization}
			
			We evaluate the performance of the linear polarization reconfigurable architecture in \cite{castellanos2023linear} in terms of the spectral efficiency and the power consumption. Here, we assume fully-digital architectures and focus on the improvements solely due to polarization diversity. As baselines, we consider a static fully-digital transmit array and a dual-polarized fully-digital transmit array. The spectral efficiency of the polarization reconfigurable architectures using the one-sided transmit polarization optimization in \cite{castellanos2023linear}. The baseband precoders for all cases are found by waterfilling over the effective channel. In the case of the dual-polarized array, note that the effective channel is unpolarized matrix on the transmit side. The dual-polarized transmit array yields the highest spectral efficiency at the cost of requiring one RF chain per antenna. The benefit of the reconfigurable array is that it increases polarization diversity with only a small increase in power consumption.

			We compute the REF as the ratio of the spectral efficiency increase to the power consumption increase.  
			Instead of picking a particular power consumption for the reconfigurable architecture, we assume that reconfiguration consumes a set fraction of the power of one RF chain. We use the model in \cite{RibeiroEtAlEnergyEfficiencyMmWaveMassive2018}, where the power consumption is divided into the power consumed by the power amplifier, $P_{\mathsf{PA}}$, the local oscillator, $P_{\mathsf{LO}}$ and each RF chain, $P_{\mathsf{RF}}$. The power consumption of the static transmit array is then
			\begin{equation}
				P_{\mathsf{cons, ST}} = P_{\mathsf{PA}} + P_{\mathsf{LO}} + \NT P_{\mathsf{RF}},
			\end{equation}
			while the power consumption of the dual-polarized transmit array is
			\begin{equation}
				P_{\mathsf{cons, DP}} = P_{\mathsf{PA}} + P_{\mathsf{LO}} + 2\NT P_{\mathsf{RF}}.
			\end{equation}
			Letting $\chi$ denote the power consumption of the reconfigurable components of an antenna normalized by the RF power consumption, the power consumption of the reconfigurable array is
			\begin{equation}
				P_{\mathsf{cons, RA}} = P_{\mathsf{PA}} + P_{\mathsf{LO}} + (1 + \chi)\NT P_{\mathsf{RF}}.
			\end{equation}
			Given a reference spectral efficiency $\mathsf{SE}_{\mathsf{ref}}$ and power consumption $P_{\mathsf{ref}}$, we denote the spectral efficiency difference as $\Delta {\mathsf{SE}} = (\mathsf{SE} - \mathsf{SE}_{\mathsf{ref}})/\mathsf{SE}_{\mathsf{ref}}$ and the power consumption difference $\Delta \Pcons = (P_{\mathsf{cons}} - P_{\mathsf{ref}})/P_{\mathsf{ref}}$. We choose the static architecture for the reference values and define the REF as 
			$\Upsilon = \frac{\Delta {\mathsf{SE}}}{\Delta \Pcons}$.
			
\begin{figure}
				\centering
				\IfFileExists{figures/pol_recon_REF.pdf}{\includegraphics[width=\linewidth]{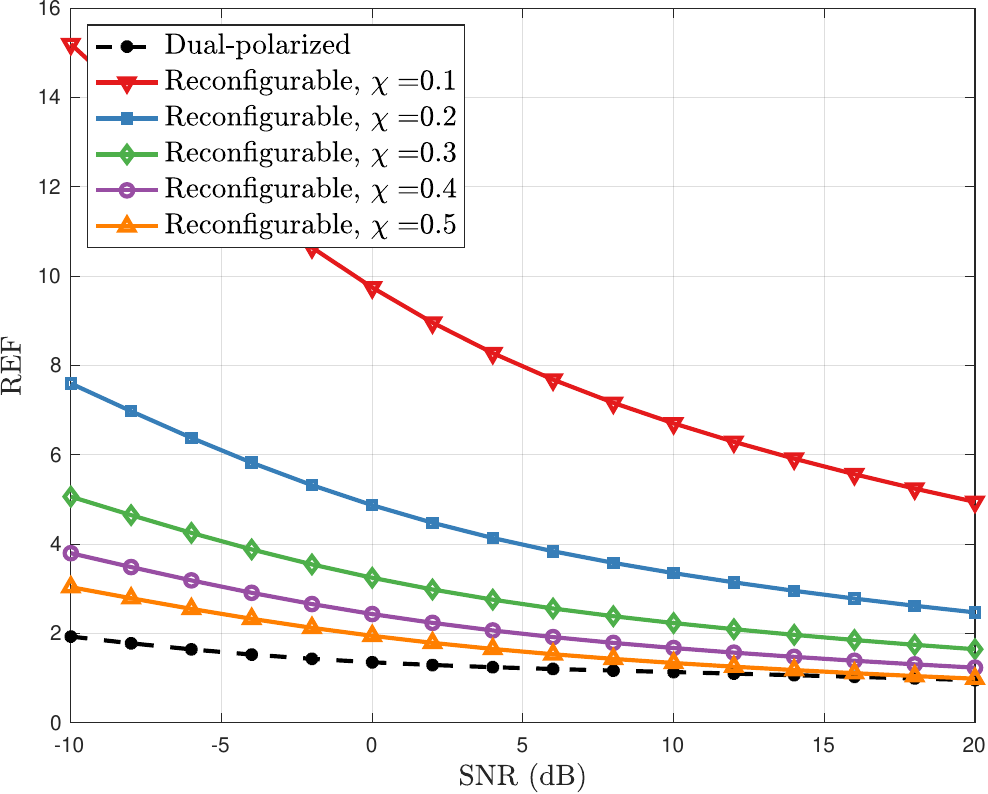}}{\fbox{\parbox[c][0.28\textheight][c]{0.92\linewidth}{\centering Polarization REF figure unavailable in the current workspace.}}}
				\caption{REF for polarization reconfigurable arrays and dual-polarized arrays in a $4 \times 4$ single-user MIMO OFDM system. The REF is defined as the ratio of the spectral efficiency increase to the power consumption increase with respect to a static transmit array. The parameter $\chi$ characterizes how much power the reconfigurable components consume compared to an RF chain. Polarization reconfigurable arrays display much higher REF values, especially at low SNR values.}
				\label{fig_pol_recon_REF}
			\end{figure}

			We show the REF of the dual-polarized array and the reconfigurable arrays as a function of SNR and $\chi$ in Fig. \ref{fig_pol_recon_REF}. We consider a single-user MIMO-OFDM system operating over $64$ subcarriers. For the single-polarization arrays, we assume uniform planar arrays with $4$ elements at both the transmitter and receiver. The dual-polarized transmit array uses $4$ pairs of orthogonally polarized antennas for a total of $8$ elements. We assume a transmit power of $P = 1$ W and the channel model in \cite{castellanos2023linear}. The transmit power determines the power amplifier power consumption through the power amplifier efficiency $\eta_{\mathsf{PA}}$ as $ P_{\mathsf{PA}} = P/\eta_{\mathsf{PA}}$. The REF of the reconfigurable array is significantly higher than that of the dual-polarized arrays since the dual-polarized array is not able to use all of the unpolarized eigenchannels through waterfilling and its main benefit is combating depolarization. The results also demonstrate that the reconfigurable arrays outperform the dual-polarized array for values of $\chi$ between 0.1 and 0.5. The results show how this type of analysis can be used to determine the performance benchmarks particular polarization reconfigurable designs should surpass to provide benefits over dual-polarized arrays.

			\section{Key takeways}\label{sec: key takeaways}
			
			This section summarizes the central insights of tri-hybrid MIMO and highlights how the integration of reconfigurable antennas reshapes signal processing, system modeling, and optimization. Unlike conventional hybrid architectures, tri-hybrid MIMO tightly couples the EM, analog, and digital domains, leading to new opportunities as well as intrinsic constraints that must be carefully addressed.
			
			\begin{itemize}
				
				\item \textbf{Tri-hybrid MIMO introduces a new signal processing paradigm:}  
				By incorporating reconfigurable antennas as a third precoding layer, the architecture moves beyond conventional hybrid MIMO, where EM behavior becomes an intrinsic component of the system model rather than a separable front-end effect.
				
				\item \textbf{Reconfigurable antennas act as EM-constrained operators:}  
				They are not merely additional precoders but physically constrained transformations that directly reshape the effective channel and radiated power. This leads to strong coupling across digital, analog, and EM layers, altering system design principles.
				
				\item \textbf{A unified input--output model enables cross-architecture understanding:}  
				The proposed formulation captures a wide range of antenna technologies (e.g., DMAs, FAS, SIM, PASS) within a common framework, allowing systematic comparison in terms of information-theoretic performance, power efficiency, and hardware constraints.
				
				\item \textbf{Hardware constraints redefine beamforming capabilities:}  
				Unlike conventional systems with near-independent per-element control, reconfigurable antennas impose structured constraints (e.g., Lorentzian coupling, discrete selection, sequential propagation), limiting beamforming flexibility and requiring new design methodologies.
				
				\item \textbf{Radiated power becomes EM-dependent:}  
				Transmit power is no longer determined solely by digital or analog precoders but depends explicitly on antenna configuration, making power modeling and optimization inherently coupled with EM behavior.
				
				\item \textbf{Optimization is inherently multi-domain and non-convex:}  
				Joint design across digital, analog, and EM layers introduces structured non-convexity, combinatorial decisions, and frequency-dependent effects, calling for new algorithmic approaches tailored to each antenna architecture.
				
				\item \textbf{REF provides a meaningful system-level performance metric:}  
				REF enables quantitative assessment of the benefits of reconfigurable antennas relative to conventional systems, revealing tradeoffs between aperture size, power efficiency, and achievable rates.
				
				\item \textbf{Tri-hybrid MIMO offers a scalable path toward large-aperture systems:}  
				By shifting part of the beamforming functionality to the EM domain, the architecture enables reduced RF chain complexity while maintaining high spatial resolution and energy efficiency.
				
			\end{itemize}
			
			Overall, tri-hybrid MIMO demonstrates that future wireless systems must move beyond purely digital abstractions and explicitly incorporate EM-domain constraints into signal processing. The key challenge, and opportunity, lies in developing unified frameworks and algorithms that fully exploit this tightly coupled multi-domain structure.

			\bibliographystyle{IEEEtran}
			\bibliography{references_merged}

		\end{document}